\documentclass[prb,aps,twocolumn,showpacs,amsmath,amssymb,floatfix,citeautoscript]{revtex4-1}

\usepackage{graphicx}
\usepackage{dcolumn}
\usepackage{amsmath}
\usepackage{amssymb}
\usepackage{times}

\begin{document}

\date{\today}
\title{Modulated Spin Liquid and Magnetic Order from a Kondo-Heisenberg model applied to $URu_{2}Si_{2}$.}

\author{X.Montiel}
\email{x.montiel1@gmail.com}
\affiliation{International Institute of Physics, Universidade Federal do Rio Grande do Norte, 59078-400 Natal-RN, Brazil\\}
\author{S.Burdin}
\email{s\'{e}bastien.burdin@u-bordeaux1.fr}
\affiliation{Univ. Bordeaux, LOMA, UMR 5798, F-33400 Talence, France}
\affiliation{CNRS, LOMA, UMR 5798, F-33400 Talence, France}
\author{C.P\'{e}pin}
\affiliation{Institut de Physique Th\'{e}orique, CEA-Saclay, 91191 Gif-sur-Yvette, France\\}
\author{A.Ferraz}
\affiliation{International Institute of Physics, Universidade Federal do Rio Grande do Norte, 59078-400 Natal-RN, Brazil\\}
\affiliation{Departamento de F\'{i}sica Te\'{o}rica e Experimental, Universidade Federal do Rio Grande do Norte, 59072-970 Natal-RN, Brazil\\}

\begin{abstract}

Within the framework of the Kondo-Heisenberg model, we analyse the effect of charge fluctuation on the modulated spin liquid (MSL) and antiferromagnetic (AF) orders which were established in a previous publication. We discuss the emergence of two quantum critical lines separating the coexisting Kondo-MSL and Kondo-MSL-AF phases. The various order parameters of the system possess a characteristic signature observable on the electronic band structure of each of the phases. We calculate that the MSL order is indeed a possible explanation of the hidden order phase of $URu_{2}Si_{2}$ heavy fermion compound. Our model produces results in qualitative agreement with the experimental (T,P) phase diagram and the two gap openings in the system and quantitative agreement with the entropy and Sommerfeld coefficient evolution.

\end{abstract}


\maketitle

\begin{section}{Introduction}
Kondo lattices are among the most studied models in condensed matter physics\cite{ACHewson}. They are realized in broad families of strongly correlated materials, when conduction electrons interact with a periodic crystal of quantum magnetic moments. 
In heavy fermion compounds, the local moments represent the $f-$electrons from the lanthanide or actinide atoms. An extremely rich variety of unusual quantum phases are observed in these systems \cite{P.Coleman3,GRStewart,VHLohneysen22}. Exotic superconductivity \cite{FSteglich1}, Quantum Critical Points (QCP) \cite{bookSachdev,QSi} and their related non-standard critical exponents result from the competition between various microscopic mechanisms \cite{Doniach,Hertz,Millis,Moriya}. Among the well accepted scenarios explaining pressure-driven QCPs in heavy fermions, the general Doniach's argument \cite{Doniach} relies on the competition between local Kondo screening and inter-moment Ruderman-Kittel-Kasuya-Yosida (RKKY) magnetic interaction\cite{Ruderman,Kasuya,Yoshida}. When the Kondo coupling dominates, the thermodynamic, transport, and magnetic properties are characterized by a Fermi liquid behavior with a large effective mass \cite{Landau, Pines,Nozieres}. The opposite regime leads generally to a magnetically ordered ground state or, alternatively, to a spin-liquid (SL) phase \cite{Anderson} when the RKKY mechanism dominates the Kondo scale but frustration prevents the magnetic ordering~\cite{S.Burdin1,BHBernhard}. Leaving aside the issue of superconductivity, it thus appears that Doniach's view \cite{Doniach} can give rise to two different kinds of QCPs. In one of them, the order parameter characterizing a {\it traditional} QCP (denoted by QCP$^c$) is the local magnetization, which can be easily measured experimentally. The second kind of QCP (denoted by QCP$^\star$) marks the breakdown of the Kondo effect, which is also called fractionalization, and has been first discussed from a theoretical ground by several authors~\cite{S.Burdin1,S.Burdin11,BHBernhard, C.Pepin2,C.Pepin3,C.Pepin2b,T.Senthil1,Hackl, QSi}. Since the volume of the Fermi surface is expected to have a significant discontinuity when crossing a QCP$^\star$ one may naively think that this single feature would be enough to provide a clear experimental signature of a Kondo breakdown transition. However, experimental reality is more complex than that, since most of the (magnetic) QCPs$^c$ also break lattice translation symmetry. This also leads to another significant variation of the Fermi surface due to the folding of the first Brillouin zone. Signatures of these quantum phase transitions can still be obtained from a Fermi surface analysis, but this requires a carefull investigation which takes into consideration the volume as well as nesting and symmetry properties.

In this article, our first aim is to use the Fermi surface analysis to clarify the signatures of the various quantum phases that emerge in Kondo lattice systems. We then apply our analysis to the specific heavy fermion compound URu$_2$Si$_2$ .

Intensive researches have been provided for almost 30 years due to its mysterious hidden order (HO) phase observed at ambient pressure below 17,5 Kelvin~\cite{JA.Mydosh}. Recent inelastic neutron scattering (INS) experiments on URu$_2$Si$_2$  \cite{A.Villaume,F.Bourdarot,CR.Wiebe,F.Bourdarot2} suggest that the HO phase is a RKKY phase somewhat similar to the antiferromagnetic (AF) phase which is realized in this compound above critical pressure. 
Indeed, neutrons, which are only sensitive to magnetism, revealed an excitation peak at a commensurate wavevector in the HO phase that coincides with the commensurate order characterizing the AF phase. This commensurate peak may be interpreted as a kind of a Bragg peak and the HO phase in this way breaks the lattice symmetry in the AF phase. 
Nevertheless, the lack (or weakness) of local magnetization in the HO phase \cite{C.Broholms,MB.Maple,Note1} establishes a clear difference with the AF order. 

Various experiments reveal the dual local and non-local nature of the wave function characterizing the hidden order \cite{JA.Mydosh}. 
The multipolar approach \cite{F.Cricchio,K.Haule,AI. Toch,H. Harima,H.Kusunose}, based on the symmetry of the Uranium orbitals, were mostly focused on a site-localized order parameter.  But these approaches, projecting the wave function onto a localized basis, do not explain the microscopic origin of the commensurate wave vector $Q_{0}$ observed by INS experiments \cite{A.Villaume,F.Bourdarot,CR.Wiebe,F.Bourdarot2}.

We recently proposed a scenario in which the HO phase is identified with a Modulated Spin Liquid (MSL) \cite{C.Pepin,C.Thomas}  with some resonant valence bonds forming a sort of singlet crystal. Our approach is based on the idea that the hidden order is characterized by a many body wave function where magnetic degrees of freedom are highly entangled from site to site. The commensurate wave vector emerges naturally from this intersite entanglement, in a similar way as discussed elsewhere \cite{H.Ideka,VP.Mineev,S.Elgazzar,AV.Balatsky,PS.Riseborough,P.Chandra}.

Introduced for the sake of clarity from a quantum Heisenberg model, the MSL has no local magnetization but it breaks the lattice translation symmetry. The AF to MSL transition is thus characterized by a melting of the staggered magnetization preserving the lattice symmetry breaking. In this MSL scenario, the partial melting of the AF order explains most of the physical properties of the AF to HO transition in URu$_2$Si$_2$. 
Yet this material is also a metallic heavy fermion and the Kondo effect needs also to be taken into acount for a correct description of its physical properties.

These considerations led us to analyse Fermi surfaces for a Kondo lattice model with an extra explicit RKKY interaction that is able to reproduce the MSL-AF transition. For the sake of clarity we consider this model on a square lattice. In section II, we present the general model and the method we use to establish the different phase diagrams presented in section III. In section IV, we apply this general model and analysis to the particular case of URu$_{2}$Si$_{2}$. An analysis of the band structure and of the Fermi surface with an emphasis on the characteristical signatures of each order parameter is presented in appendice.

\end{section}
\begin{section}{Model and method}

\subsection{Model}
We consider a Kondo-Heisenberg model \cite{SenguptaGeorges,IgLacoq} on a square lattice with $N$ sites 
and a lattice constant $a=1$, defined by the following Hamiltonian: 
\begin{eqnarray}
\label{11}
H &=&H_{KL}+H_{RKKY} \nonumber\\
&~&\nonumber\\
&\equiv&t_{c}\sum_{\langle R,R' \rangle,\alpha}c_{R\alpha}^{\dag}c_{R'\alpha} 
+J_{K}\sum_{R}{\bf S}_{R}\cdot{\bf s}_{R}
+J\sum_{\langle R,R'\rangle}{\bf S}_{R}\cdot{\bf S}_{R'}~,  \nonumber\\
&~&
\end{eqnarray} 
where the operator $c_{R\alpha}^{(\dag)}$ annihilates (creates) a conduction electron on site $R$ with spin 
component $\alpha=\uparrow,\downarrow\equiv\pm 1$, and ${\bf S}_R$ denotes a quantum spin $1/2$ on site $R$. The sum over $\langle R,R'\rangle$ refers to nearest neighbors with each bond being counted only once. The conduction electron local spin density can be expressed as  ${\bf s}_{R}=\frac{1}{2}\sum_{\alpha\beta}c^{\dag}_{R\alpha}\boldsymbol{\sigma}_{\alpha\beta}c_{R\beta}$, where $\boldsymbol{\sigma}\equiv (\sigma^x,\sigma^y,\sigma^z)$ denotes the Pauli matrices. The average electronic occupation per site is fixed to be $n_{c}$, which will later be taken into acount by the introduction of a given chemical potential $\mu$. Here, $H_{KL}$ is a Kondo lattice Hamiltonian, with a nearest neighbor hopping term $t_{c}$ and a local Kondo antiferromagnetic coupling $J_K$ between conduction electrons and local moments. The Heisenberg term $H_{RKKY}$ adds an antiferromagnetic interaction $J$ between nearest neighboring Kondo spins. 

Hereafter, we use the Abrikosov pseudofermions representation for the local spin $1/2$ operators \cite{AAAbrikosov}: 
${\bf S}_{R}=\frac{1}{2}\sum_{\alpha\beta}f^{\dag}_{R\alpha}\boldsymbol{\sigma}_{\alpha\beta}f_{R\beta}$, where the
$f_{R\alpha}$ ($f_{R\alpha}^{\dag}$) are fermionic annihilation (creation) operators satisfying the local single-particule occupation constraint: 
\begin{eqnarray}
\label{Fermionicconstraint}
f_{R\uparrow}^{\dag}f_{R\uparrow}+f_{R\downarrow}^{\dag}f_{R\downarrow}=1~. 
\end{eqnarray}
The Kondo interaction is first rewritten as 
$J_K{\bf S}_{R}\cdot{\bf s}_{R}=\frac{J_K}{2}\left( 
\sum_{\alpha}f_{R\alpha}^{\dag}c_{R\alpha}\right)\left( 
\sum_{\beta}f_{R\beta}c_{R\beta}^{\dag}\right) -J_Kn_c/4$. 
These local terms are then decoupled using a standard mean-field approximation~\cite{C.Lacroix,Coleman83,Reads}, within the Hubbard Stratonovitch scheme. The RKKY interaction is treated with exactly the same mean-field procedure as the one described in Ref.~\cite{C.Pepin}: the Heisenberg interaction is decoupled on each nearest neighbor bond $\langle R,R'\rangle$, partially in a spin-liquid channel, and partially in an antiferromagnetic Weiss field channel. The respective weights, $J_{SL}/J$ and $J_{AF}/J$, of each decoupling channel are constrained by the relation $J\equiv J_{SL}+J_{AF}$. Here, $J_{SL}$ and $J_{AF}$ can be considered as tuning parameters for the model. This decoupling scheme, which might appear to be arbitrary on a square lattice model, captures phenomenologicaly some of the frustration effects of a more realistic three-dimensional model~\cite{C.Thomas,CThomas2}. 

\subsection{Mean-field method}
We introduce the mean field decouplings in the Kondo, the Weiss-AF, and the SL channels, considering only the colinear order for the AF channel. The Hamiltonian~(\ref{11}) is approximated as: 
\begin{eqnarray}
\label{12}
H&\approx& H_{MF}\equiv\sum_{\alpha}\left( H^{\alpha}_{KL}+H^{\alpha}_{RKKY}\right)+E_0\nonumber\\
&&+\mu\sum_{R,\alpha}\left( \frac{n_c}{2}-c_{R\alpha}^{\dag}c_{R\alpha}\right) 
+\lambda\sum_{R,\alpha}\left( \frac{1}{2}-f_{R\alpha}^{\dag}f_{R\alpha}\right)~, \nonumber\\
&&~
\end{eqnarray}
with
\begin{eqnarray}
H^{\alpha}_{KL}&=&t_{c}\sum_{ \langle R,R'\rangle}c_{R\alpha}^{\dag}c_{R'\alpha}
+\sum_{R}(b_{R}c_{R\alpha}^{\dag}f_{R\alpha}+h.c.) \nonumber\\
H^{\alpha}_{RKKY}&=&\sum_{ \langle R,R'\rangle}(\varphi_{RR'}f_{R\alpha}^{\dag}f_{R'\alpha}+h.c.)\nonumber\\
&&+\alpha\sum_{R}f_{R\alpha}^{\dag}f_{R\alpha}
\sum_z m_{R+z}\nonumber \\
E_{0}&=&\sum_{R}\frac{|b_{R}|^{2}}{J_{K}}
+\sum_{\langle R,R'\rangle}\left(\frac{|\varphi_{RR'}|^{2}}{J_{SL}}-\frac{m_{R}m_{R'}}{2J_{AF}}\right)~, 
\end{eqnarray}
where $\lambda$ is a Lagrange multiplier introduced in order to take into acount the local constraints Eq.~(\ref{Fermionicconstraint}). Following the standard mean-field scheme~\cite{C.Lacroix,Coleman83,Reads}, this Lagrange multiplier field has been assumed to be constant and homogeneous. Therefore, $\lambda$ acts here as an effective chemical potential for the Abrikosov fermions and the constraints~(\ref{Fermionicconstraint}) are satisfied on average only. The sum over z denotes a sum over all the nearest neighbour sites. The Kondo mean-field parameter is given by the self consistent equation $b_{R}=J_{K}\sum_{\alpha}\langle f_{R\alpha}^{\dag} c_{R\alpha}\rangle$, where the thermal average $\langle\cdots\rangle$ is computed from the mean-field hamiltonian~(\ref{12}). This Kondo parameter can be understood as an effective hybridization between the light conduction electron band and the heavy flat band~\cite{Kondo,C.Lacroix,Coleman83,Reads}.  The spin liquid and antiferromagnetic mean-field self-consistent relations give $\varphi_{RR'}=-J_{SL}\sum_{\alpha}\langle f_{R\alpha}^{\dag}f_{R'\alpha}\rangle$ and $m_{R}=J_{AF}\sum_{\alpha}\langle \alpha f_{R\alpha}^{\dag}f_{R\alpha}\rangle$. A full self-consistent resolution of the mean-field effective model also requires the determination of $\mu$ and $\lambda$ from the relations $\frac{1}{N}\sum_{R,\alpha}{\langle c_{R\alpha}^{\dag}c_{R\alpha}\rangle}=n_{c}$ and $\frac{1}{N}\sum_{R,\alpha}{\langle f_{R\alpha}^{\dag}f_{R\alpha}\rangle}=1$. 

In principle, the Kondo and the Weiss fields $b_R$ and $m_R$ are site dependent whereas the SL field $\varphi_{RR'}$ is bond dependent instead. Following the procedure of Refs.\cite{C.Pepin,C.Thomas}, we will make some simplifying ansatz for the RKKY fields. First, we remark that the AF field $m_{R}$ is formally similar to the standard Weiss mean-field moment that would emerge from a classical approximation. Since we consider a square lattice model, we will consider only the N\'eel ordering to represent the AF state. We thus assume that $m_{R}=S_Q~e^{iQ\cdot R}$, where $Q=(\pi,\pi)$ and $S_Q$ is the staggered magnetization. The SL field $\varphi_{RR'}$ is reminiscent of the Resonant Valence Bond (RVB) state introduced by Anderson and co-workers\cite{Anderson, Baskaran}. The RVB state was originaly introduced within an homogeneous SL field. This was motivated by the description of frustrated spin systems that could not form a long range magnetically ordered state. Nevertheless, it was shown in Ref~\cite{C.Pepin} that a spatially modulated spin liquid (MSL) was more energetically stable on a square lattice than an homogeneous SL phase. Here, we assume a real space modulation of the SL mean field which has the form $\varphi_{RR'}=\delta_{RR'}{\phi_{0}+i \frac{\phi_{Q}}{2}\sum_{\pm}{e^{\pm iQ.(R+R')/2}}}$. The SL field is defined on the dual (i.e., bond) lattice in general, although this particular MSL ansatz breaks the same lattice translation symmetry which is broken by the AF N\'eel ordering. Consequently, the Bravais lattices for the MSL and AF ordered state are identical to each other and correspond to a doubled unit cell in contrast with the single unit of the initial square lattice. We make an ansatz for the Kondo parameter $b_R$. An indirect coupling between the modulated spin-liquid order parameter and the Kondo screening may produce a modulation of the Kondo parameter as well. Thus, we include the possibility of a modulation for the Kondo mean field parameter writing $b_{R}=b_{0}+ib_{Q}e^{iQ.R}$, where both $b_0$ and $b_Q$ are real. With this ansatz, the phase of the Kondo mean-field parameter can be spatially modulated, when $b_Q\neq 0$. But its amplitude $\rho\equiv\sqrt{ b_0^2+b_Q^2}$ is homogeneous since $e^{iQ.R}$ reduces only to $\pm 1$. Note that the amplitudet $\rho$ could also be modulated. Indeed, a space modulation of $\rho$ may lead to charge ordering. We left this possibility aside because it has never been observed in the URu$_2$Si$_2$ compound which is the main motivation of this work. Finally, the Kondo parameters are reexpressed as $b_{0}=\rho\cos(\theta)$ and $b_{Q}=\rho\sin(\theta)$, with a non-zero value of the amplitude $\rho$ being typical for a Kondo phase. A non-zero angle $\theta$ clearly indicates that the Kondo phase is modulated. We define the Kondo temperature $T_{K}$ as the temperature at which the conduction electrons start to screen the local moments. Numerically, we signalled this temperature as the temperature where the amplitude $\rho$ becomes different from zero.

Invoking the Fourier transform, $f_{k\alpha}\equiv\frac{1}{\sqrt{N}}\sum_Re^{^{ikR}}f_{R\alpha}$ and its inverse, and replacing the site and bond dependent mean-fields by their corresponding ansatz expressions, the effective mean-field Hamiltonian~(\ref{12}) reads:
\begin{eqnarray}
\label{Mean-field-Hamiltonian}
H_{MF}&=&\sum_{k,\alpha}\left[ 
\left(t_{c}\gamma_{k}-\mu\right)c_{k\alpha}^{\dag}c_{k\alpha}
+\left(\phi_{0}\gamma_{k} -\lambda\right)f_{k\alpha}^{\dag}f_{k\alpha}\right]
\nonumber \\
&&+\sum_{k,\alpha}\left(4\alpha S_{Q}+i\phi_{Q}\gamma_{k-Q/2}\right)
f_{k\alpha}^{\dag}f_{k-Q\alpha}\nonumber \\
&&+\rho\cos(\theta)\sum_{k,\alpha}(c_{k\alpha}^{\dag}f_{k\alpha}+f_{k\alpha}^{\dag}c_{k\alpha}) \nonumber \\
&&+i\rho\sin(\theta)\sum_{k,\alpha}(f_{k-Q\alpha}^{\dag}c_{k\alpha}-c_{k-Q\alpha}^{\dag}f_{k\alpha}) \nonumber \\
&&+N\left[\frac{\rho^{2}}{J_{K}}+\frac{2\left(\phi_{0}^{2}+\phi_{Q}^{2}\right)}{J_{SL}}+\frac{S_{Q}^{2}}{J_{AF}}
+\lambda +\mu n_{c} \right]~,
\end{eqnarray}
where $\gamma_{k}\equiv -2\cos(k_{x})-2\cos(k_{y})$ is the square lattice dispersion. 

The mean-field parameters $\phi_{0}$, $\phi_{Q}$, $S_{Q}$, $\rho$, $\theta$, the Lagrange multiplier $\lambda$, and the chemical potential $\mu$, are determined self-consistently by the minimization of the free energy, given by $\beta{\cal F}_{MF}=-Tr[\exp{(-\beta H_{MF})}]$. We find the following saddle point equations: 

\begin{eqnarray}
\phi_0&=&-\frac{J_{SL}}{4N}\sum_{k,\alpha}\gamma_k\langle f_{k\alpha}^{\dag}f_{k\alpha}\rangle~, 
\label{EqMF-phi0}
\end{eqnarray}
\begin{eqnarray}
\phi_Q&=&
-\frac{iJ_{SL}}{4N}\sum_{k,\alpha}\gamma_{k-Q/2}\langle f_{k\alpha}^{\dag}f_{k-Q\alpha}\rangle~, 
\label{EqMF-phiQ}
\end{eqnarray}
\begin{eqnarray}
S_Q&=&-\frac{2J_{AF}}{N}\sum_{k,\alpha}\alpha\langle f_{k\alpha}^{\dag}f_{k\alpha}\rangle~, 
\label{EqMF-SQ}
\end{eqnarray}
\begin{eqnarray}
\rho\cos(\theta)&=&-\frac{J_{K}}{2N}\sum_{k,\alpha}\langle c_{k\alpha}^{\dag}f_{k\alpha}
+h.c.\rangle~, \label{EqMF-rhocos}
\end{eqnarray}
\begin{eqnarray}
\rho\sin(\theta)&=&-\frac{J_{K}}{2N}\sum_{k,\alpha}\langle i f_{k-Q\alpha}^{\dag}c_{k\alpha}
+h.c.\rangle~, \label{EqMF-rhosin}
\end{eqnarray}
\begin{eqnarray}
1&=&\frac{1}{N}\sum_{k,\alpha}\langle f_{k\alpha}^{\dag}f_{k\alpha}\rangle~, \label{EqMF-lambda}
\end{eqnarray}
\begin{eqnarray}
n_c&=&\frac{1}{N}\sum_{k,\alpha}\langle c_{k\alpha}^{\dag}c_{k\alpha}\rangle~. \label{EqMF-mu}
\end{eqnarray}

\subsection{Numerical treatment of the mean-field equations}
Within the mean field Hamiltonian, the free energy per site can be written explicitly as a function of the mean-field parameters $\phi_{0}$, $\phi_{Q}$, $S_{Q}$, $\rho$, $\theta$, the Lagrange multiplier $\lambda$, and the chemical potential $\mu$. We find: 
\begin{eqnarray}
\label{110} 
{\cal F}_{MF}&=&-\frac{k_{B}T}{N}\sum_{k,n}ln\left(1+e^{\left(-\beta\Omega_{n}\right)}\right)+\lambda+\mu n_{c} \nonumber \\
&&+\frac{\rho^{2}}{J_{K}}+\frac{2\left(\phi_{0}^{2}+\phi_{Q}^{2}\right)}{J_{SL}}+\frac{S_{Q}^{2}}{J_{AF}}
\end{eqnarray}
 where $\beta=1/k_{B}T$ , $N$ is the number of site, $ n$  runs over the band defined by the eigenvalues of the effective mean field Hamiltonian and $\Omega_{n}$ are the eigenvalues of the following mean field Hamiltonian matrix 

\begin{widetext}
\begin{equation} \left (
\begin{array}{cccc}
t_{c} \gamma_{k}-\mu&0&\rho cos(\theta)& -i\rho sin(\theta)\\
0&t_{c} \gamma_{k-Q}-\mu&-i\rho sin(\theta)&\rho cos(\theta)\\
\rho cos(\theta)& i\rho sin(\theta)&\phi_{0}\gamma_{k}-\lambda&4 \alpha S_{Q}+i \phi_{Q} \gamma_{k-\frac{Q}{2}}\\
i\rho sin(\theta)&\rho cos(\theta)&4 \alpha S_{Q}-i\phi_{Q}\gamma_{k-\frac{Q}{2}}&\phi_{0}\gamma_{k-Q}-\lambda
\end{array}\right )
\end{equation}
\end{widetext}

For a numerical self-consistent solution, we proceed iteratively as follows. At each loop, for a fixed $\left(\phi_{0},\phi_{Q},S_{Q},r_{0},r_{Q}\right)$, the chemical potentials $\lambda$ and $\mu$ are determined by dichotomy method to fullfill simultaneously the constraint conditions (\ref{EqMF-lambda}) and (\ref{EqMF-mu}). In turn, for either fixed $\lambda$ and $\mu$, the free energy expression~(\ref{110}) is minimized using Powell´s method~\cite{Powell}, providing the solution $\left(\phi_{0},\phi_{Q},S_{Q},\rho,\theta\right)$ of the mean-field equations.~(\ref{EqMF-phi0}, \ref{EqMF-phiQ}, \ref{EqMF-SQ}, \ref{EqMF-rhocos}, \ref{EqMF-rhosin} ).

All the numerical results presented in this article have been obtained with the following choice of parameteres: the Kondo coupling is set to be $J_{K}/t_{c} = 0.8686$ where $t_{c}$  is the conduction electron hopping. $t_{c}$  characterizes the conduction electron bandwidth which equals $8t_{c}$ in a square lattice. These parameters are choosen such that the Kondo temperature, $T_{K0}$, for a zero RKKY interaction ($J_{AF}=J_{SL}=0$) is small compared to the bandwidth ($T_{K0}=0.098t_{c}$). This guaranties that we work in the weak Kondo coupling regime $T_{K} \ll t_{c}$. We choose also an electronic filling $n_{c}=0.7$ which avoids the square lattice instability (at $n_{c}=0.5$)  and which also avoids the Kondo insulating regime (at $n_{c}=1$). We used other sets of numerical values for testing, that are not presented here, and we obtained the same qualitative conclusions as the ones presented in this article.

\end{section}

\begin{section}{Phase diagram}
In this section we analyse the phase diagram obtained for the Kondo-Heisenberg model within the mean-field approximation. 
Considering the different mean-field parameters introduced in the previous section, $\phi_{0}$, $\phi_{Q}$, $S_{Q}$, $\rho$, and $\theta$, we study the competition or the cooperation between Kondo screening, and AF or Modulated SL ordering. 
Our very systematic approach starts from an arbitrary fixed value of electronic filling $n_c$. The pure Kondo lattice part of the Hamiltonian~(\ref{11}), $H_{KL}$, is thus roughly characterized by one energy scale \cite{BGG2000} : the non-interacting Kondo temperature $T_{K0}$. Doniach's argument~\cite{Doniach} can be reproduced phenomenologicaly within a Kondo-Heisenberg model by comparing $T_{K0}$ with the energy scale 
that characterizes the Heisenberg part $H_{RKKY}$ of the Hamiltonian. Therefore, in this section we fix the $T_{K0}$ that provides the energy unit scale of the problem, and we analyse the phase diagram as a function of the two RKKY phenomenological energies $J_{AF}$ and $J_{SL}$, and the temperature $T$. The resulting phase diagram is depicted schematically in figure~\ref{fig99}, on the basis of the numerical solution of the self-consistent mean-field relations, that is shown in figure~\ref{fig85}. 

\begin{subsection}{Description of phases and order parameters \label{Section-definition-phases}}

As expected, a Kondo regime is obtained at low temperature for sufficiently small RKKY energies. In contrast pure RKKY-dominated phases appear when $J_{AF}$ or $J_{SL}$ are large enough. 
The general phase diagram (see Figure~\ref{fig99}) reveals also coexisting phases: for the Kondo and the MSL order (KMSL) and for the Kondo phase, MSL order and AF order (KMSL-AF). Note that the AF order can coexist with Kondo effect only in the presence of MSL order. The phase diagram reveals also a QCP$^\star$ and a QCP$^c$ at $J_{SL}^{\star}$, $J_{SL}^{c}$ , respectively, and $J_{AF}^{\star c}$ which will be discussed further. 
Among the five mean-field parameters that have been introduced, $\phi_{Q}$, $S_{Q}$, and $\theta$ correspond to true order parameters characterizing phase transitions, whilst $\phi_{0}$ and $\rho$ rather characterize crossovers to 
correlated regimes, which are associated respectively with the spin-liquid and the Kondo heavy-fermion phases. 
Hereafter, we classify these phases in four sets: the disordered phase, the pure AF, the pure MSL, and the mixed AF+MSL 
ordered phases. 

\begin{figure}
  \includegraphics{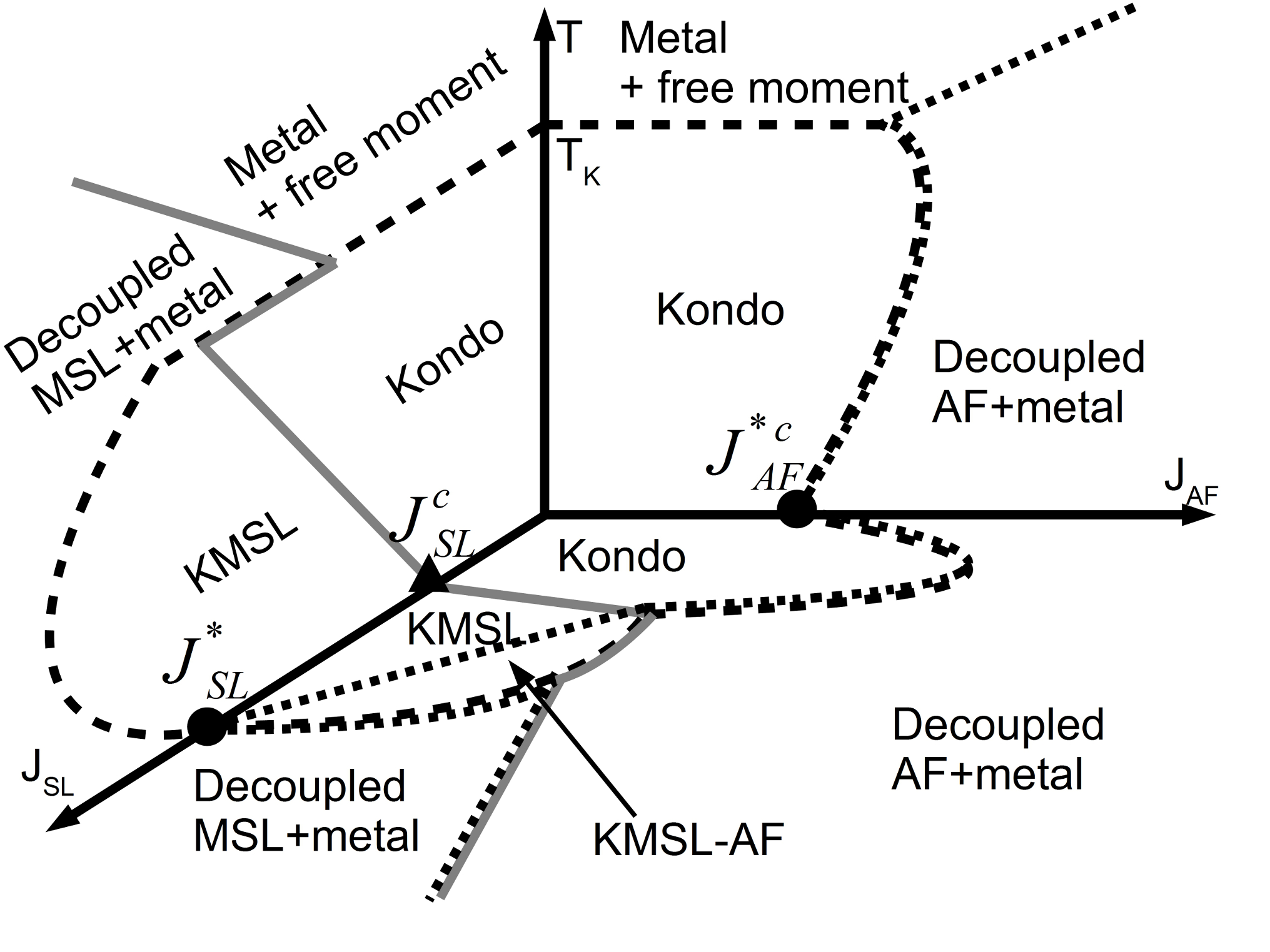}
  \caption{ \label{fig99} Schematic phase diagram of the Kondo-Heisenberg model in coordinates $(T,J_{AF},J_{SL})$, 
showing: the MSL phase transition (solid line), the AF phase transition (doted line), and the Kondo crossover (dashed line). 
Three specific QCP are marked: two of them indicate a breakdown of Kondo effect (circle) and the third one indicates the 
onset of MSL (triangle) in a Kondo regime. The various phases and their related order parameters are defined in 
section~\ref{Section-definition-phases}.}
\end{figure}

\begin{subsubsection}{The Disordered phase}
The disordered phase covers the paramagnetic high temperature metallic and the Kondo correlated heavy fermion phases. 
Within our mean-field approach, the crossover Kondo temperature is signalled by a transition at $T_K$ which is 
characterized by a continuous vanishing of the Kondo parameter: for $T<T_K$ with the Kondo regime being realized for $\rho \neq 0$. 
Below $T_{K}$, the mean-field approach may distinguish two kinds of Kondo regimes, depending on the value of the homogeneous SL parameter $\phi_0$. The Kondo regime with a finite $\phi_{0}$ does not break any symmetry. 
\begin{paragraph}{The decoupled regime}
~~\\
This regime is established for a temperature higher than all the interaction energy scales, i.e., when 
$T>T_{K0}, J_{SL}, J_{AF}$. At the mean-field level, this corresponds to a complete decoupling between the Kondo spins and the conduction electrons, with $\phi_{0}=\phi_{Q}=S_{Q}=\rho=0$. 
The transport and thermodynamic properties correspond to that of a light metal, and the resulting magnetic susceptibility can be directly associated with free moments. 
\end{paragraph}
\begin{paragraph}{The Kondo regime with $\phi_{0}=0$}
~~\\
This usual Kondo regime is defined by the condition $\rho\neq 0$, with all other mean-field parameters being nullified. 
It is obtained for $T<T_{K}$ when $J_{SL} \approx 0$, and it extends to $J_{AF}<J_{AF}^{\star c}$ 
as shown in figure~\ref{fig85}. 
\end{paragraph}
\begin{paragraph}{The Kondo regime with $\phi_{0} \neq 0$}
~~\\
Finite values of both $\phi_{0}$ and $\rho$ characterize this correlated heavy-fermion regime which does not break any symmetry ($\phi_{Q}=S_{Q}=0$). 
Note that the solutions $\theta=-\pi/2$ with $\phi_{0} > 0$ and $\theta=0$ with $\phi_{0}<0$ are equivalent to each other. 
The apparent symmetry breaking provided by $\theta=-\pi/2$ can be recovered by the local gauge transformation $(\phi_{0}>0,\theta=-\pi/2) \rightarrow (\phi_{0}<0,\theta=0)$. Physically, this corresponds to a Kondo coupling between conduction electrons and local moments dispersion with opposite sign.

\end{paragraph}
\end{subsubsection}

\begin{subsubsection}{The pure AF phase}
The AF phase is obtained at low temperature for $J_{AF}>J_{AF}^{\star c}$. Here, the occurance of long range magnetic AF ordering coincides with the suppression of the Kondo effect. This decoupled AF+metal phase is characterized within our mean-field 
approach, by a finite staggered magnetization $S_{Q}$  together with $\phi_{0}=\phi_{Q}=\rho=0$. 
We find no coexistence between Kondo effect and AF order, but such coexisting phases might appear in different 
lattice structures.  
The pure AF phase breaks lattice and time-reversal symmetries. 
\end{subsubsection}

\begin{subsubsection}{The MSL phases}
The MSL order parameter merges with the mean-field $\phi_{Q}$ which was introduced in \cite{C.Pepin,C.Thomas} within purely Heisenberg models. Here, the presence of conduction electrons and their Kondo coupling to the local moments 
lead to two MSL regimes:  
\begin{paragraph}{The pure, decoupled, MSL phase}
~~\\
This phase is purely RKKY-dominated phase, characterized by a finite MSL order parameter $\phi_{Q}$, without 
Kondo effect ($\rho=0$), and without AF order ($S_{Q}= 0$). Here, the conduction electrons are effectively decoupled from the Kondo spins which form the MSL order. The lattice translation symmetry is broken but the local magnetization remains equal to zero. The Z4 lattice symmetry is also broken as discussed in Ref.~\cite{C.Pepin}. 
\end{paragraph}

\begin{paragraph}{The Kondo-MSL phase}
~~\\
The KMSL phase merges at low temperature for a broad range of intermediate couplings $J_{SL}^{c}<J_{SL}<J_{SL}^{\star}$. In this phase, conduction electrons are strongly correlated to the MSL ordering formed by the local moments. Within our mean-field approach, this phase is defined by a finite Kondo parameter $\rho$ coexisting with a finite MSL order parameter $\phi_{Q}$. Here, the order is purely MSL, which means that the system has no staggered magnetization, i.e. $S_Q=0$. We find that the Kondo effective hybridization is inhomogeneous, with $\theta=-\pi/2$. 

Such a phase modulation in the Kondo parameter results from the lattice breaking of symmetry of the MSL. It is still compatible with the local $U(1)$ gauge invariance invoked in our mean-field approach. Indeed, without the MSL order, any local phase variation of the Kondo hybridization could be gauged out invoking a local $U(1)$ transformation. However, in the presence of an MSL order, such a local gauge transformation for the Kondo term would also have some effect on the intersite spin-liquid term. Using appropriate local gauge transformations, one might thus be able to map a set of mean-field solution $\theta=-\pi/2$ to $\theta=0$ but at the same time $\phi_0$ and $\phi_Q$ must be mapped to $-\phi_{0}$ and $-\phi_{Q}$ respectively

This strongly correlated phase breaks the square lattice translation and the Z4 symmetries.
\end{paragraph}
\end{subsubsection}

\begin{subsubsection}{The mixed AF+MSL phase}
\begin{paragraph}{The decoupled AF+MSL phase}
~~\\
When $J_{AF}$ and $J_{SL}$ are of similar magnitude and are both much larger than $T_{K0}$, the system forms 
an RKKY-dominated phase with conduction electrons effectively decoupled from the local moments. This 
AF+MSL decoupled phase is defined by finite $ S_Q$ and $\phi_{Q}$, together with $\rho=0$. This appears in 
a very narrow part of the phase diagram. Such a phase breaks time reversal symmetry as well as the lattice translation and Z4 symmetries. 
\end{paragraph}
\begin{paragraph}{The Kondo coupled AF+MSL phase}
~~\\
This Kondo coupled AF and MSL phase (KMSL-AF) is characterized by non-zero values of all the considered mean-field parameters, 
$\phi_{Q}$, $\phi_{0}$, $S_{Q}$, and $\rho$. It is realized when the three relevent energy scales, $T_{K0}$, 
$J_{AF}$, and $J_{SL}$ are of the same order of magnitude. This phase breaks the same symmetries broken by the decoupled AF+MSL phase. 
\end{paragraph}
\end{subsubsection}

\end{subsection}

\begin{subsection}{Numerical results}

\begin{figure}
  \includegraphics{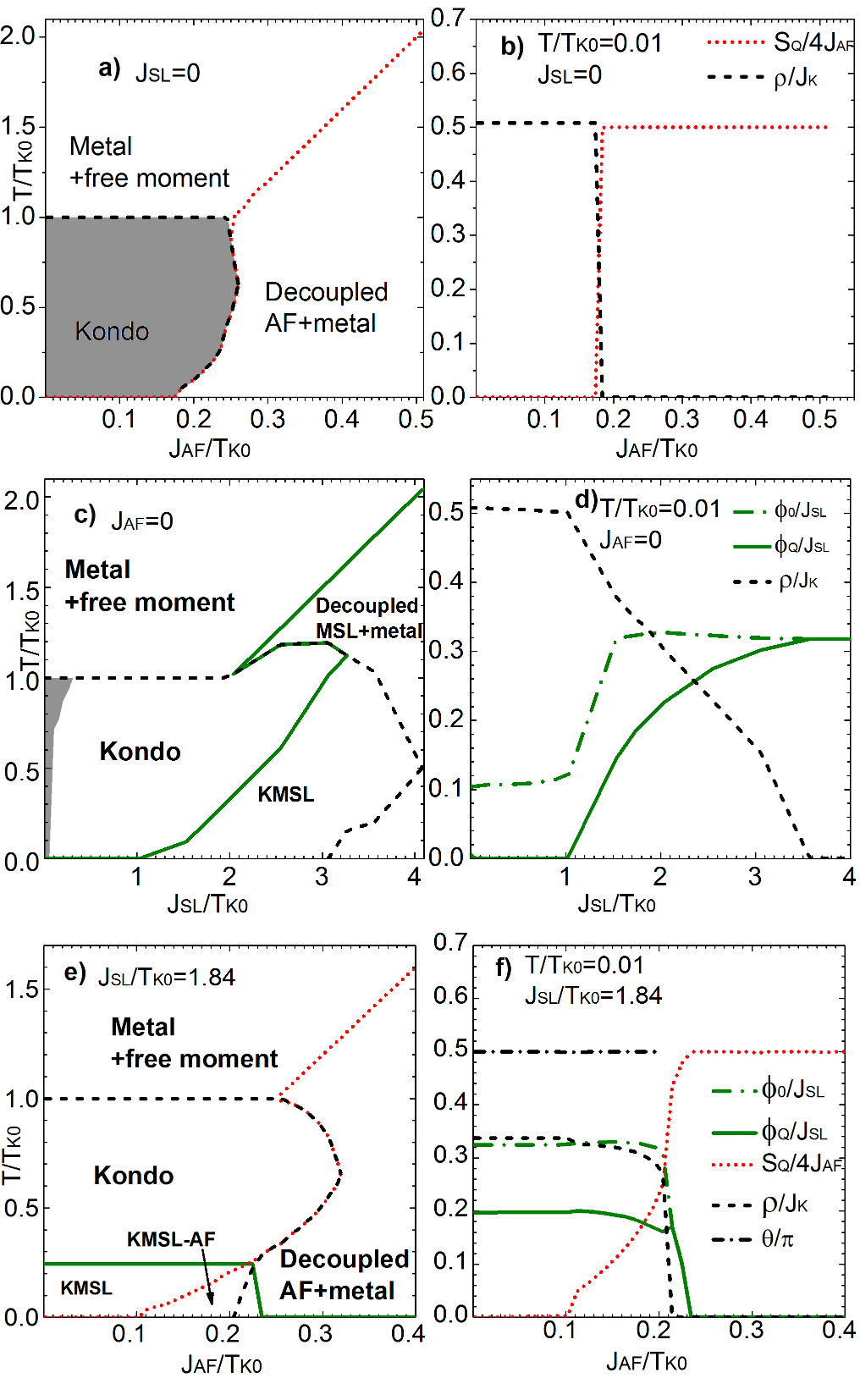}
  \caption{ \label{fig85} (Color online) Phase diagrams ( a) c) e)) and variation of the mean fields at $T/T_{K0}=0.01$ ( b) d) f))  for $J_{K}/t_{c}=0.8686$, the conducting electron density $n_{c}=0.7$ and $T_{K0}/t_{c}=0.098$  where the MSL phase transition is presented in solid line, the AF phase transition in doted line, and the Kondo crossover in dashed line. We chose the regime $J_{SL}=0$ as a function of $J_{AF}$ in a) and b). We show the regime $J_{AF}=0$ as a function of $J_{SL}$  in c) and d) and an intermediary case $J_{SL}^{c}<J_{SL}<J_{SL}^{\star}$  as a function of $J_{AF}$ in e) and f). The grey area delimit the Kondo phase with $\phi_{0}=0$.}
\end{figure}

For $J_{SL}=0$, the Kondo and AF order competition exhibits a QCP at $J^{\star c}_{AF}=0.2T_{K0}$ (see figure\ref{fig85}.a ) which is both a QCP$^{\star}$ (breaking of the Kondo effect) and a QCP$^{c}$ (onset of long range magnetic order). The value of $J^{\star c}_{AF}$ is of the same order of $T_{K0}$ and this is consistent with the Doniach's arguments. The Kondo to AF transition is of first order nature as emphasized  by the discontinuities of the effective Kondo coupling $\rho$ and the staggered magnetization field $S_{Q}$ at the transition (see figure\ref{fig85}.b ). The Kondo to Normal metal phase transition manifests itself as a second order phase transition because of the mean field approximation but it should indeed be a crossover. The AF to metal phase transition is second order in our case. 

For $J_{AF}=0$, both the Kondo local and spin liquid intersite screening reveals two QCPs (see figure\ref{fig85}.c) : a QCP$^{c}$ at $J_{SL}=T_{K0}$ and a QCP$^{\star}$  at $J_{SL}^{\star}=3T_{K0}$.
The QCP$^{c}$ at $J_{SL}=T_{K0}$ characterizes a second order phase transition from the Kondo phase with $\phi_{0} \neq 0$ to the KMSL phase hybrid phase as shown by the continuity of the modulated spin liquid field $\phi_{Q}$ at the transition (see figure\ref{fig85}.d).
The QCP$^\star$ at $J_{SL}=3T_{K0}$ appears at the KMSL to MSL second order phase transition as emphasized by the continuous vanishing ot the effective Kondo field  $\rho$ at the transition. In the MSL phase, the homogeneous and modulated spin liquid mean fields $\phi_{0}$ and $\phi_{Q}$  have the same values $\phi_{0}=\phi_{Q}$. This is related to the absence of second nearest neighbour coupling and was first signalized in \cite{C.Pepin}. The metal to Kondo and Kondo with $\phi_{0} =0$ to Kondo with $\phi_{0} \neq 0$ changes of state appears as phase transition because of the mean field treatment but are both crossovers. The metal to MSL phase transition is second order. In contrast, the transition from the Kondo with $\phi=0$ state to a MSL phase is a first order phase transition.  

The KMSL-AF phase appears at an intermediary SL coupling $J^{c}_{SL}<J_{SL}<J^{c}_{SL}$ as shown in the figures  \ref{fig99} and \ref{fig85}.e. In the KMSL-AF phase, the staggered magnetisation $S_{Q}$ increases with the AF coupling as shown in figure \ref{fig85}.f. The phase transition between the KMSL and the KMSL-AF phases is second order and this is signaled by the continuous appearence of $S_{Q}$ in figure \ref{fig85} f. However, the transition from the KMSL-AF to the MSL+AF phases and from the MSL+AF to AF phases are both first order as attested by the discontinuity of the order parameters $\phi_{0}$, $\phi_{Q}$ and $S_{Q}$ mean fields at the transitions (see figure\ref{fig85}.f ). We note the existence of the MSL+AF coexisting phase in a very narrow part of the phase diagram (between the KMSL-AF and AF phases). This coexisting phase disappears for larger values of $J_{SL}$ and temperature. The $\theta$ phase is equals to $-\pi/2$ in the KMSL and KMSL-AF phases and vanishes with the effective Kondo mean field $\rho$.
The metal to Kondo phase change is a crossover while the kondo to KMSL and the metal to AF changes of state are real second order phase transitions. Finally, the Kondo to AF change is a first order phase transition.

\end{subsection}

\begin{subsection}{Discussion}
From this model, we show the emergence of two quantum critical lines in the (JAF, JSL) phase diagramm, ending on three QCPs when either JAF or JSL vanishes. One line, denoted with a $\star$ marks the Kondo breakdown. The other critical line, with a $c$ index, characterizes the breaking of lattice translation symmetry.  
. A QCP$^{\star c}$ appears at the Kondo-AF phase transition $(J_{SL}=0,J_{AF}=0.2T_{K0})$ and a QCP$^{\star}$  occurs at the KMSL to MSL phase transition for $(J_{SL}=3T_{K0}, J_{AF}=0)$.  A QCP$^{c}$ separates the Kondo and KMSL phases  $(J_{SL}=T_{K0},J_{AF}=0)$.

A QCP$^\star$ separates a heavy Fermi liquid phase where the conventionnal Luttinger theorem \cite{Abrikosovbook,Luttinger1,Luttinger2} holds, and a fractionalized phase in which such a theorem does not apply any longer. 
In other words, the number of quasiparticles in the heavy Fermi liquid phase is the sum of the conduction electrons and localized electrons. Contrary to that, light conduction electrons provide the metallic properties of the fractionalized phase. 
From this theoretical definition emerged the following experimental signature of a Kondo breakdown QCP$^\star$ in a periodic system: the Fermi surface is expected to vary from a large to a small volume as the QCP$^\star$ is crossed from the Kondo to the RKKY phase. Other signatures of a QCP$^\star$ are also expected for crystals as well as for non periodic systems, including, for example, a change of sign in the Hall constant. This mechanism drives the physics in the vicinities of both the QCP$^{\star}$ ($J_{SL}=3T_{K0}$) and the QCP$^{\star c}$ ($J_{AF}=0.2T_{K0}$).

When the RKKY interaction produces a magnetically ordered phase, the magnetic QCP$^c$ might coincide or not with the Kondo breakdown  QCP$^\star$. This issue relies on the possible coexistence of the Kondo effect with magnetic ordering since QCP$^c$=QCP$^\star$ means that the local Kondo screening disappears precisely when the magnetic moments order sets in. 
This issue depends on several model parameters including the nature of magnetic ordering, and on the dimensionality of the physical system \cite{Isaev}. This mechanism occurs close to the  QCP$^{c}$ ($J_{SL}=T_{K0}$) and the QCP$^{\star c}$ ($J_{AF}=0.2T_{K0}$).
The physics near the QCP$^{\star c}$ is driven by the Kondo breakdown and lattice symmetry breaking.
\end{subsection}

\end{section}

\begin{section}{Hidden order phase as modulated spin liquid  in $URu_{2}Si_{2}$}
In this section, we apply the formalism developped in the section II, III and IV to the particular case of $URu_{2}Si_{2}$.  

The MSL model assumes that HO order parameter has both local and itinerant characteristics \cite{C.Pepin,C.Thomas}. The MSL model reproduces qualitatively different experimental observations such as the entropy quenching, the (T,P) phase diagram and the MSL gap evolution. 

In the following, the Kondo coupling, $J_{K}=54.37meV$ and the conduction electron hopping $t_{c}=63.31meV$ are set to establish a Kondo temperature around $T_{K}=70K$. This value of Kondo temperature is suggested by resistivity measurements \cite{TTM.Palstra,MB.Maple,E.Hassinger} and optical conductivity \cite{RLobo} experiments. Hall experiments \cite{Schoenes,YSOh,YKasahara} and Angle Resolved Photo-Emission Spectroscopy (ARPES) \cite{SantanderSyro,RYoshida,SChatterjee,SantanderSyro2} measurments confirm the existence of hybridized bands.

The spin liquid coupling $J_{SL}=11.37meV$ is adjusted to the HO phase transition temperature $T_{0}=17.5K$  as indicated in the specific heat data \cite{TTM.Palstra}. The tunning of $J_{AF}$ reproduces the pressure tunning given by the experiments \cite{C.Thomas}.

\begin{subsection}{The variation of entropy and specific heat}

\begin{subsubsection}{Main Experimental results}
The hidden order phase is signalled by a large jump in the specific heat at $17.5K$ \cite{TTM.Palstra,JA.Mydosh} characterizing a second order phase transition. The total entropy quenched in this phase is around $0.3k_{B}ln(2)$ per U atoms \cite{TTM.Palstra,MB.Maple}. In spite of this entropy quenching, the measured magnetic moment is $m= 0.03.\mu_{B}$ \cite{C.Broholms}. Consequently, the HO phase cannot be explained by an antiferromagnetic (AF) phase \cite{MB.Maple}.
\end{subsubsection}

\begin{subsubsection}{Numerical results}
\begin{figure}
  \includegraphics{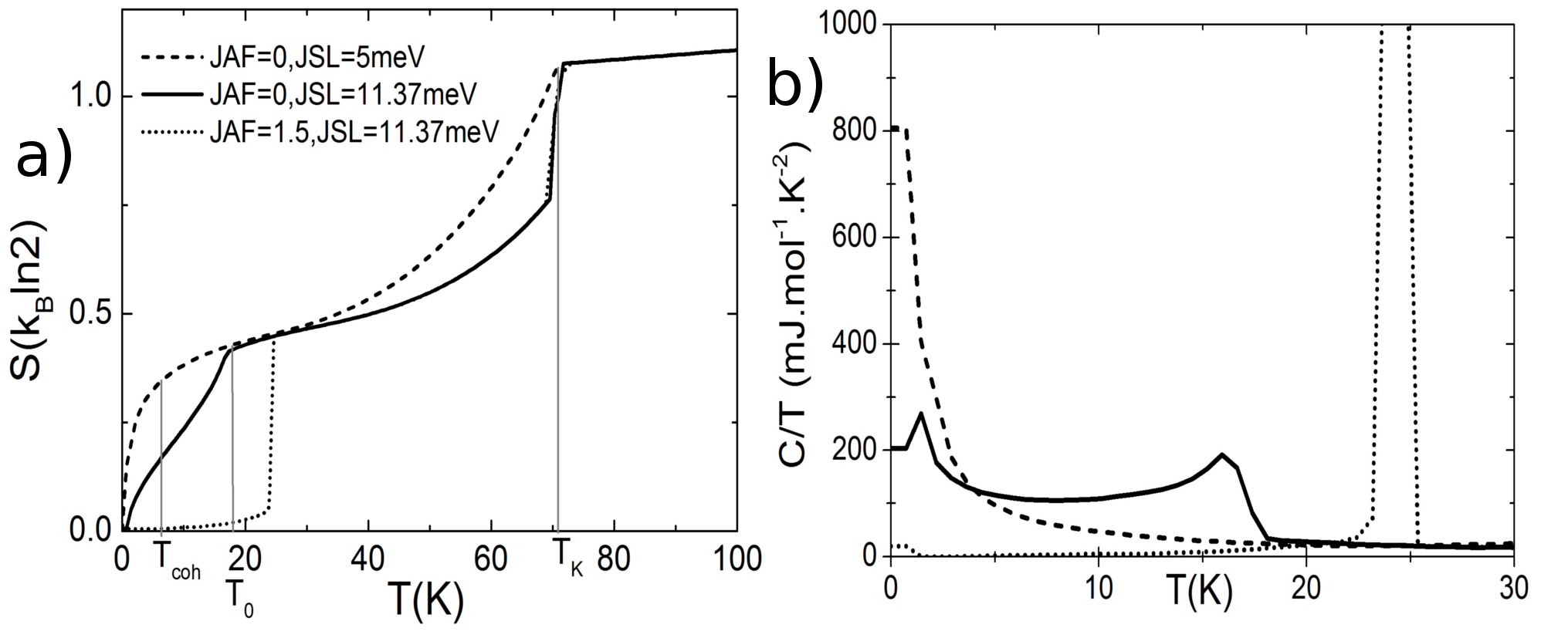}
  \caption{ \label{fig14} Temperature dependence of (a) the entropy,  (b) the Sommerfeld coefficient C/T, for a pure Kondo ground state (dashed line, $J_{SL}$=5 meV and $J_{AF}$=0), a Kondo-MSL ground state (solid line, $J_{SL}$=11.37 meV and $J_{AF}$=0), and an AF ground state (doted line, $J_{SL}$=11.37 meV and $J_{AF}$=1.5 meV). Note that the entropy evolution around TK is abrupt because of the mean-field approximation for the Kondo interaction. A crossover is expected around TK beyond the mean-field. We clearly see that the Kondo-MSL scenario, unlike the pure Kondo one, is characterized by a peak in C/T around T=17K. Note that the too-sharp peak obtained at the N\'eel temperature within the AF scenario (doted line) should be broaden within a more realistic three-dimensional lattice (see the text).}
\end{figure}

The entropy S per site, with $S=-\partial {\cal F}_{MF}/\partial T$, is plotted in figure \ref{fig14} a) for three following cases : a pure Kondo phase with $T_{K}=70K$ (dashed line), a Kondo + $\phi_{0}$  phase with $T_{K}=70K$ and MSL order with $T_{0}=17,5K$ (solid line) and a Kondo + $\phi_{0}$  phase with $T_{K}=70K$ and AF order with $T_{N}=24K$ (dotted line). 

The entropy is quenched around the Kondo temperature $T_{K}\approx 70K$ giving $S=k_{B}ln2$. Note that the little difference in entropy with $k_{B}ln2$ at $T_{K}$ in figure $\ref{fig14}$ originates from the entropy of the conduction electron that are not all frozen at $T_{K}$. The abrupt decrease of entropy around $T=T_{K}$ is an artifact of the mean field approximation that describes the Kondo crossover by a phase transition. The entropy in the Kondo phase below $T_{K}$ presents two different regimes of decreasing entropy (dashed line on figure \ref{fig14} a) ). The first regime occurs around $T_{K}$ and is related to the local Kondo screening. The second regime, occuring for $T<T_{coh}$ , is related to the coherence of the Fermi liquid. This smooth crossover at $T_{coh}$ manifests itself with the appearence of the coherent Fermi liquid. In this regime, the entropy decreases linearly to zero. This behaviour is coherent with earlier results (see e.g. Ref \cite{BGG2000}).

The entropy for higher spin liquid coupling $J_{SL}=11.37meV$ is represented in figure \ref{fig14} a) by the solid and dotted lines. The variaiton of the entropy between the pure Kondo and Kondo with $\phi_{0}$ phases (respectively dashed and solid (and dotted) lines on figure \ref{fig14} a) ) can be explained by the spin liquid correlations appearing with the Kondo screening at $T_{K}$  that contributes to freeze a part of that entropy. Note that in the model presented here, we do not take into account the orbital symmetry of the Uranium atoms and the crystal field splitting that also contributes to the determination of $T_{K}$.

At the Kondo to KMSL phase transition, the variation of entropy takes place at $T_{0}$ (see dashed line in figure \ref{fig14} a) ). This change can be related to a jump in specific heat as observed in real compound \cite{TTM.Palstra,E.Hassinger}.  Moreover, the variation of entropy between $T_{0}$ and $T=0K$ is around $0.3k_{B}ln2$  which is near the value measured in the bulk coumpound \cite{TTM.Palstra,MB.Maple}. This variation is also seen at the Kondo-AF transition (dotted line on the figure \ref{fig14} a) ) and this suggests a similar mechanism for this entropy quench between KMSL and AF phases.

Note that the coherence temperature of the Kondo lattice $T_{coh}$ is smaller than the KMSL phase transition temperature $T_{0}$. This implies that the Kondo screening is not complete at $T_{0}$. The partial Kondo coupling existing above $T_{0}$ may explain the band structure reconfiguration observed around $30K$ in both optical conductivity \cite{DABonn,JLevallois}, ARPES experiments \cite{SChatterjee} and Scanning Tunneling Microscopy (STM) \cite{PAynajian}. It is a strong feature of our model that it describes the formation of the Kondo screening within a simple coupling to an itinerant band.

The Sommerfeld coefficient $C/T=\partial S/\partial T$ is presented in figure \ref{fig14} b). We see a jump in the specific heat at the Kondo to KMSL transition (solid line) and at the Kondo to AF transition (dotted line) corresponding to a second order phase transition. Note that the jump in specific heat appears at $C/T \approx 200mJ mol^{-1} K^{-2}$ which is close to the experimental value. From the three curve of the figure \ref{fig14} b), we show that Kondo lattice Fermi liquid coherence only cannot  reproduce a peak in C/T around 17K. In order to reproduce this peak, a RKKY-like mechanism is required and this is provided here by the MSL ordering.  

The value of the peak of the Sommerfeld coefficient at the AF phase transition is higher than in the real compound. This results from the square lattice approximation in which the Kondo breakdown coincides with AF order. We are aware that a 3-dimensional model might lead to a coexistence between Kondo and AF; in that case we could expect the peak in C/T to be less sharp at the AF transition.  Note that the peak in Sommerfeld coefficient appearing below $T_{0}$ around $T \approx T_{coh}$ characterizes the formation of the coherent state constituting the Fermi liquid. This peak does not appear in the real compound where a signature of a superconducting state is observed. 

\end{subsubsection}
\end{subsection}

\begin{subsection}{The (T,P) phase diagram and evolution of Fermi surface}
\begin{subsubsection}{Main Experimental results}
The Temperature-Pressure phase diagram was extracted from neutron scattering data \cite{H.Amitsuka,A.Villaume,F.Bourdarot} and from resistivity measurements \cite{E.Hassinger}. It  shows a second order HO phase transition at $T=17,5K$ at ambient pressure. However, it has been shown that HO phase suffers a first order phase transition to AF phase with magnetic moment $m=\mu_{B}$ at $P=0.5GPa$ \cite{H.Amitsuka,E.Hassinger,A.Villaume,F.Bourdarot}. 
Conductivity measurements \cite{MB.Maple,JRJeffries},angle resolved photoemission spectroscopy (ARPES) \cite{SantanderSyro}, infrared spectroscopy \cite{DABonn,JLevallois}, Hall effect \cite{Schoenes,YSOh} and STM \cite{PAynajian} show a gap opening at the Fermi surface inducing  a Fermi surface reconstruction at the HO phase transition. Hence Fermi surface and transport properties exhibit strong similarities in both the HO and the AF phases \cite{EHassinger2}.
\end{subsubsection}
\begin{subsubsection}{Numerical results}
\begin{figure}
  \includegraphics{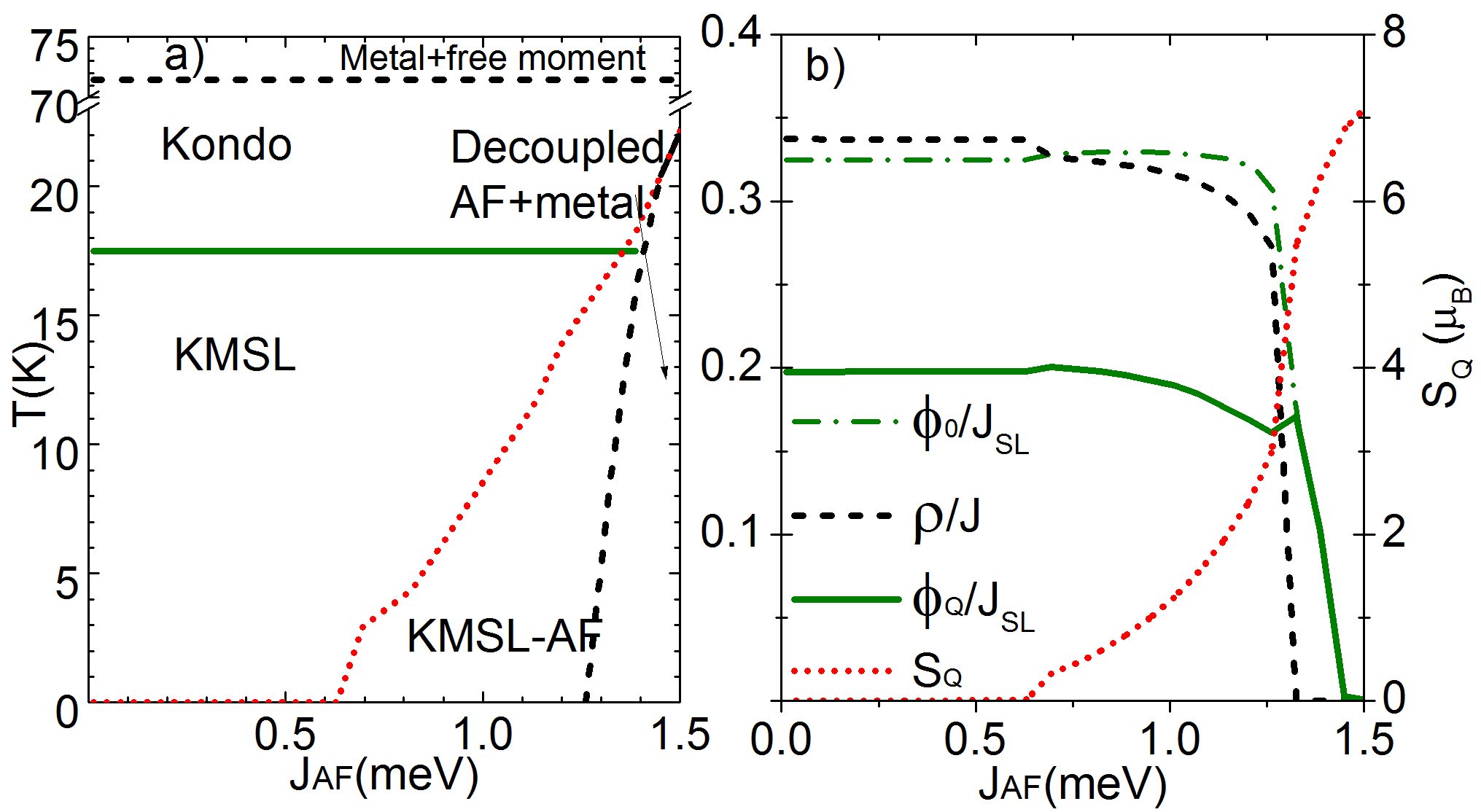}
  \caption{ \label{fig12}  a) $(T,J_{AF})$ phase diagram in $URu_{2}Si_{2}$ with $J_{SL}=11,37meV$, $J_{K}=54,37meV$ and $t_{c}=63,31meV$. Here $J_{AF}$ represents phenomenologically the effect of pressure. b) Evolution of the mean field parameter and the staggered magnetization $S_{Q}$ for the ground state. In both of figure, $J_{SL}=11,37meV$, $J_{K}=54,37meV$ and $t_{c}=63,31meV$.}
\end{figure}
\begin{figure}
  \includegraphics{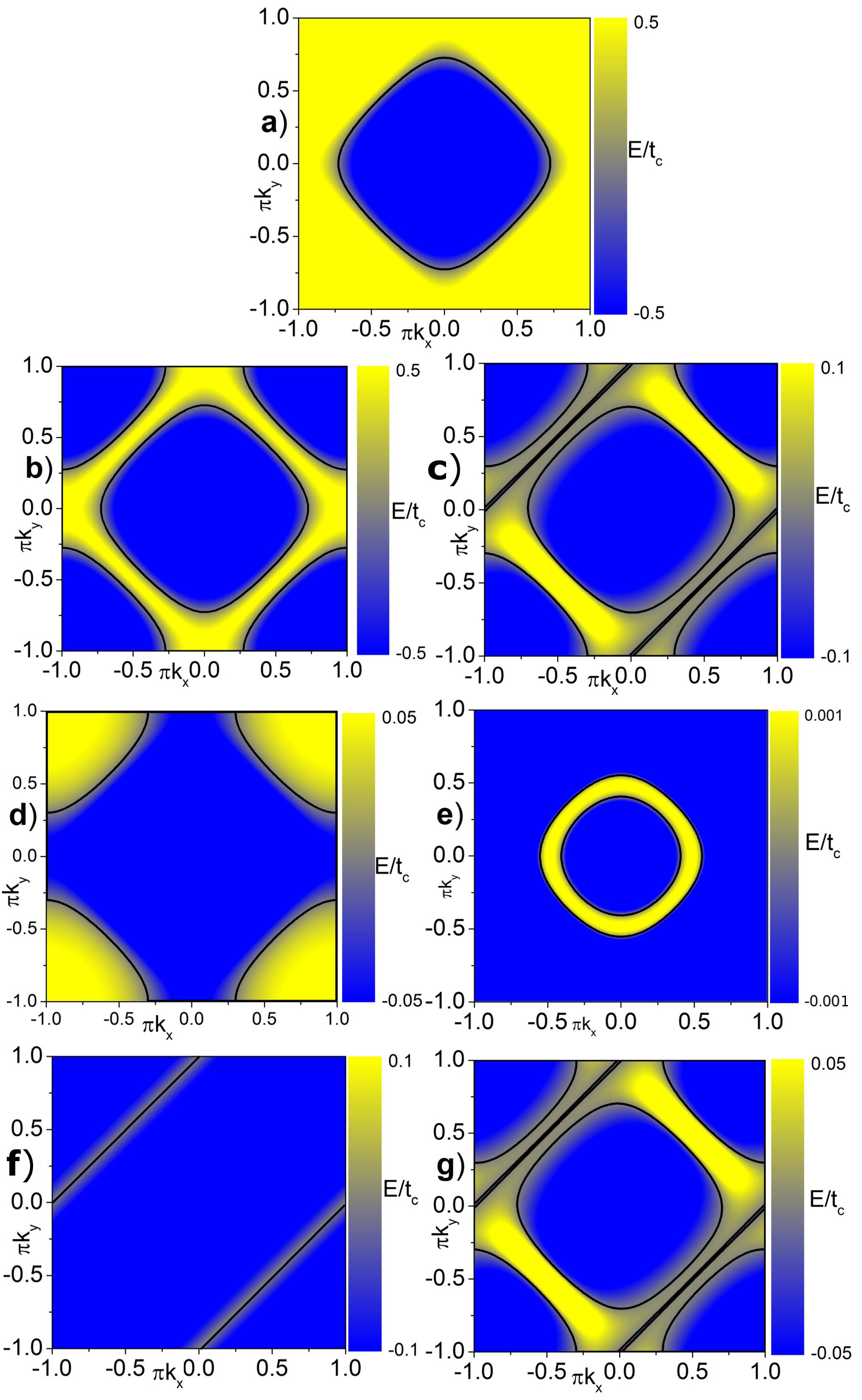}
  \caption{ \label{fig122} Dispersions of a) the paramagnetic metal phase, b) the AF phase, c) the KMSL phase, d) the Kondo phase with $\phi=0$,and with e) $\phi \neq 0$, f) the spinons of the MSL phase and g) the KMSL-AF phase in the first Brillouin zone. The solid line represents the Fermi surface at the Fermi level ($E=0$). In e), the hole states are confined between the two solid lines. The similarity between the AF (Fig. b) and MSL (Fig. c) folded Fermi surfaces is consistent with quantum oscillation experiments realized under pressure $\cite{EHassinger2}$.}
\end{figure}

The $(T,J_{AF})$ phase diagram presented in figure \ref{fig12} shows a second order phase transition between the Kondo and the KMSL phases and a first order transition between the KMSL-AF and AF phases. This is in agreement with the INS experiments  \cite{H.Amitsuka,A.Villaume,F.Bourdarot}.
We observe a strong Fermi surface reconstruction with folding at the Kondo-KMSL phase transition (at $T_{0}=17.5K$) (see figure \ref{fig122} e) and c) ). Note that in our model, this reconstruction is associated with the Z4 and lattice symmetry breaking appearing simultaneously with the MSL order parameter. 
As is displayed by quantum oscillation experiments \cite{EHassinger2}, the HO and AF phases exhibits similar Fermi surfaces. This similarity results from a folding of the initial Fermi surface (see figure \ref{fig122} a), b) and c)). Moreover, in our model, these similarity is emphasized by the progressive appearence of AF order in KMSL phase that steadily destroys the Z4 symmetry breaking (see figure \ref{fig122} b), c) and g)). A detailed analysis of the Fermi surfaces presented on the figure \ref{fig122} is proposed in the appendice B.
\end{subsubsection}
\end{subsection}

\begin{subsection}{Evolution of MSL gap}
\begin{subsubsection}{Main Experimental results}
Inelastic neutron scattering experiments exhibit a resonance at the commensurate wave vector $Q_{0}=\frac{2\pi}{a}(1, 0, 0)$ which transforms itself into a strong elastic AF signal for a pressure $P > 5kbar$ \cite{CR.Wiebe,F.Bourdarot2}. An inelastic resonance occurs in the AF and HO phases at the incommensurate wave vector $Q^{*} = \frac{2\pi}{a}(1\pm 0.4, 0, 0)$. Two distincts gaps exist in the system : $\Delta_{Q^{*}} \approx 5meV$ which is related to the transport properties and $\Delta_{Q_{0}}\approx2.5meV$ which characterizes the HO phase itself \cite{CR.Wiebe,F.Bourdarot2}.
\end{subsubsection}
\begin{subsubsection}{Numerical results}
\begin{figure}
  \includegraphics{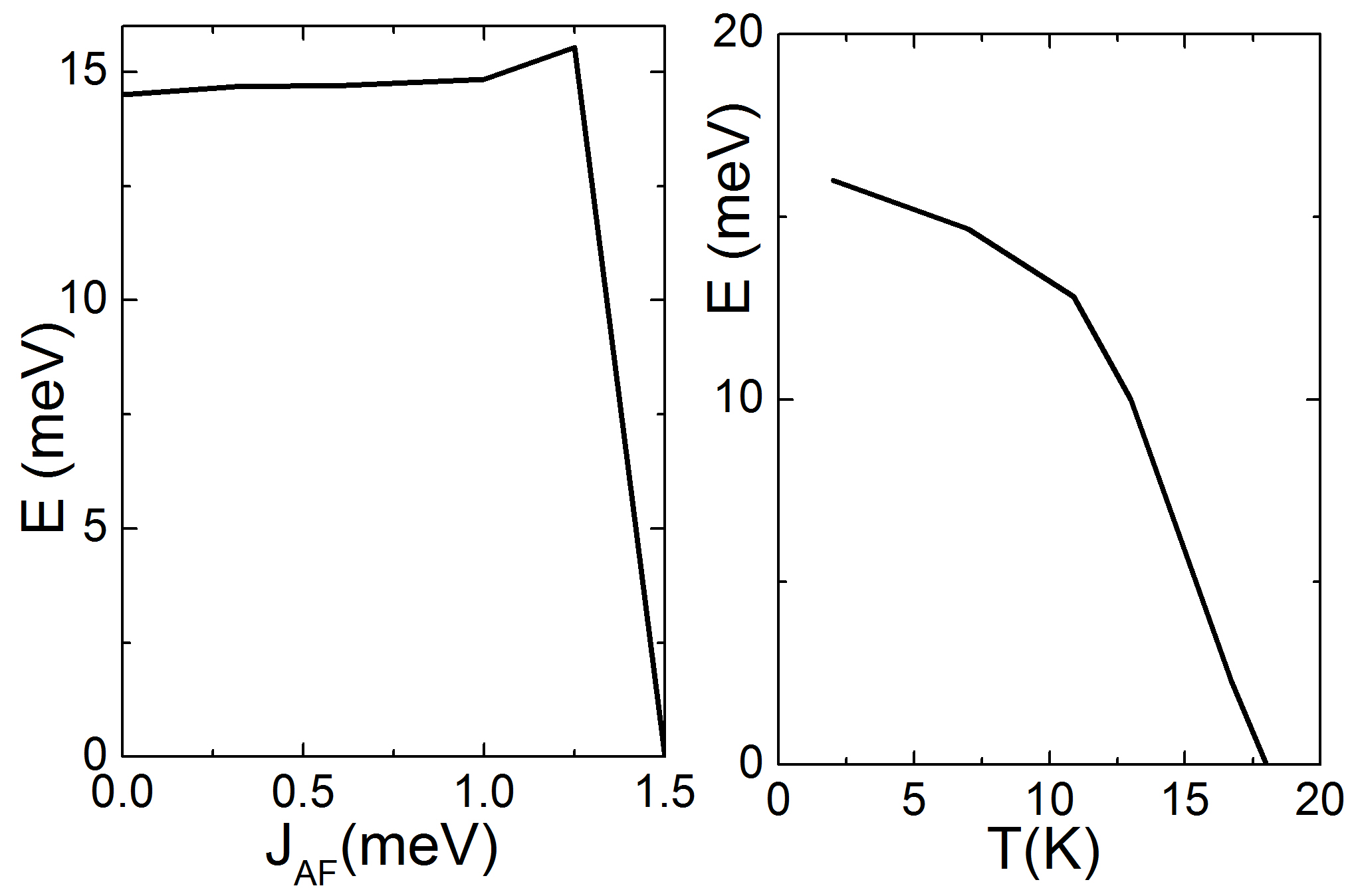}
  \caption{ \label{fig13} a) Direct gap at the $X_{1}$ point as a function of $J_{AF}$ for ground state. b)  Direct gap at the $X_{1}$ point as a function of the temperature in KMSL state ($J_{AF}=0.31 meV$).}
\end{figure}

The direct gap, $\Delta_{X_{1}}$ (see in Appendice A Introduction, section 1 c and 1.d) exists only in the KMSL state and vanishes at the KMSL-AF to AF phase transition (see figure \ref{fig13}). Moreover, $\Delta_{X_{1}}$ is associated with the vector $Q$ (see in Appendice A section 1.c and 1.d). 
 $\Delta_{X_{1}}$ vanishes continuously at the temperature $T=17,5K=T_{0}$ confirming the second order nature of the KMSL phase transition. 
\end{subsubsection}
\end{subsection}

\begin{subsection}{Evolution of the spectral density}
In this section we analyse the local density of states $\rho(\omega)=\frac{-1}{2N\pi}\sum_{k}{Im(G_c(\omega, k))}$ where the conduction electron Green function $G_c(\omega, k)$ is computed from the mean field Hamiltonian (\ref{Mean-field-Hamiltonian}).
\begin{subsubsection}{Main Experimental results}
A gap appearing around T=30-80K have been detected by optical conductivity measurments \cite{DABonn,JLevallois}. The conclusions of the authors is that the band structure reconstruction around T=30K is a precursor to the HO phase transition T=17,5K. These observations have been emphasized by the observation of a band close to the Fermi level by ARPES \cite{SChatterjee} and pseudo-gap measurments by STM \cite{PAynajian}.
\end{subsubsection}

\begin{subsubsection}{Numerical results}
\begin{figure}
  \includegraphics{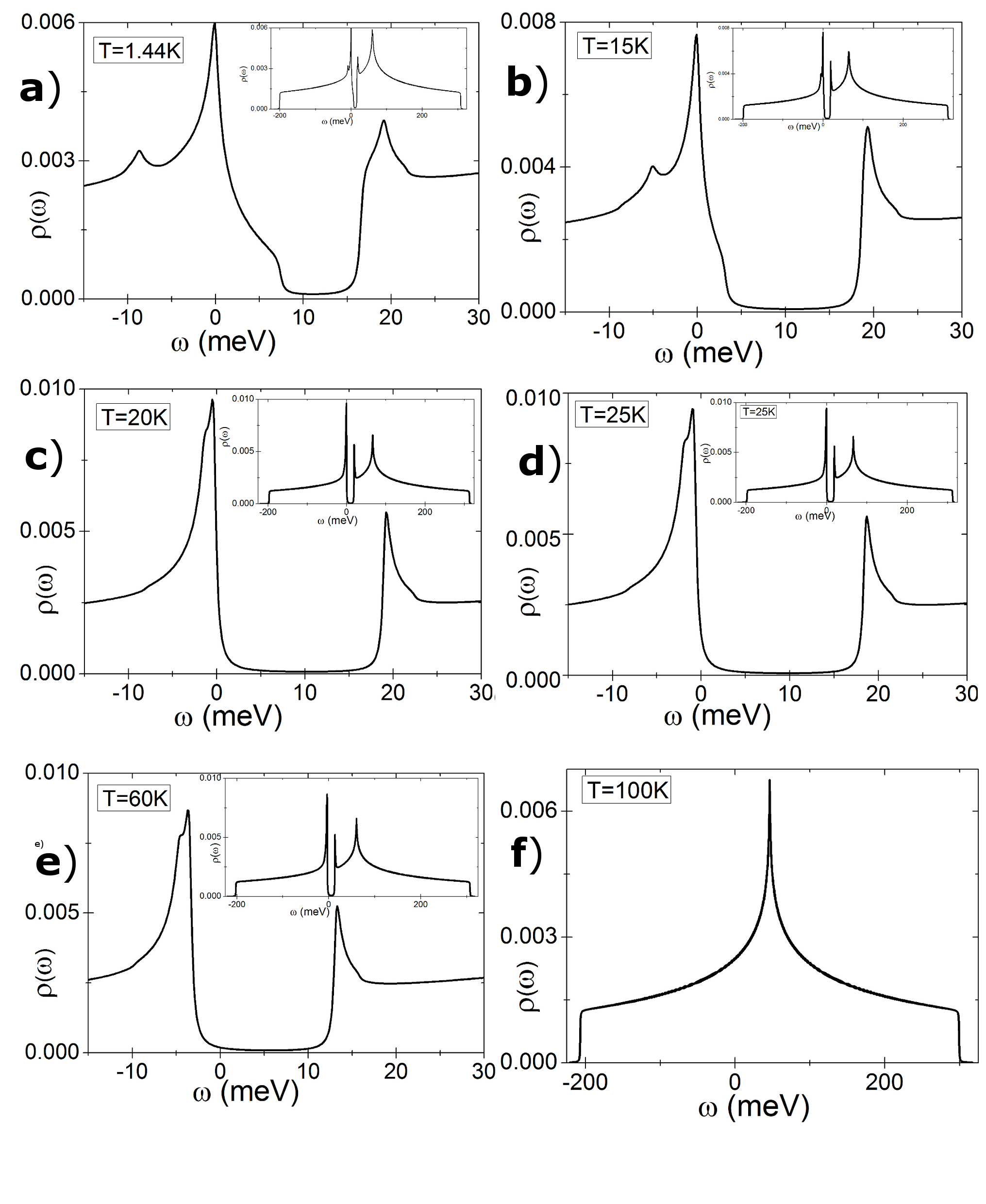}
  \caption{ \label{fig133} Spectral density in the $URu_{2}Si_{2}$ for a KMSL ground state ($J_{SL}=11.37meV$ and $J_{AF}=0$) in the effectively decoupled high temperature phase a) at T=100K, the Kondo phase b) at T=60K, c) at T=25K and d) at T=20K and in the KMSL phase e) at T=15K and f) at T=1.44K. The van Hove singularity appearing in figure a) is standard for tight-binding model on a square lattice, and it would coincide with the Fermi level only at electronic half-filling. Here, we precisely chose nc=0.7 in order to locate the Fermi level sufficiently away from this singularity which has no physical meaning for $URu_{2}Si_{2}$.}
\end{figure}
The evolution of the spectral density for different temperature is presented in figure \ref{fig133}. We see that a gap appears in the Kondo phase which increases with the decrease of temperature. This gap has a value close to 20meV for T=25K. This value is closed to the experimental value (12meV) \cite{DABonn,JLevallois}. Below the KMSL phase transition, we see the appearence of an additionnal low energy structure which is characteristic of the MSL order. We think that the gap observed above $T_{0}$ in the optical conductivity measurements is directly correlated to the partial Kondo screening occuring at this temperature. This gap remain independant of the KMSL gap which appear below $T_{0}$. 
Note that the gap observed in optical conductivity and the gap associated with the incommensurate wave vector $Q^{\star}$, $\Delta_{Q^{\star}}$ observed by neutron scattering are both related to the transport properties of the compound.
\end{subsubsection}
\end{subsection}

\end{section}

\begin{section}{Conclusion}
 We demonstrate the existence of a stable solution for the Kondo-MSL-AF competition and generalize earlier approach \cite{C.Pepin,C.Thomas} taking into account the conducting regime. The modulated spin liquid is a non conventional magnetic ordered phase which  competes with the Kondo screening in a similar way as to what happens with the AF magnetic order. The existence of two QCPs that emerged from the competition of Kondo and MSL is the signature of our model. In particular, the KMSL phase, which sees the coexistence of Kondo and MSL order, is a new outcome of our model.

This stable solution produces two distinct gaps in the electronic band structure. One gap is related to the Kondo screening as well as to the transport properties. The other gap is related to the MSL order parameter and the commensurate vector $Q$. These two gaps can be considered independently and mark a clear distinction between the magnetic and the conducting properties of these systems. The existence of these two gaps on the square lattice is another promising feature of this model.

 Our results are in qualitative good agreement with the experimental phase diagram for the $URu_{2}Si_{2}$ compound \cite{A.Villaume}. We found a first order KMSL to AF phase transition and a second order Kondo to KMSL  phase transition. Futhermore, the evolution of the Fermi surfaces denotes a Fermi surface reconstruction at the KSL/KMSL phase transition \cite{SantanderSyro,RYoshida,SChatterjee,SantanderSyro2}. Note that in our model, the reconstruction of the Fermi surface at the Kondo-KMSL phase transition is related to the MSL order parameter unlike what is proposed by other workers in \cite{DABonn,JLevallois,SChatterjee} . We also find a variation of entropy around $0.3Rln(2)$ closed to the variation of entropy measured in the realistic compound \cite{TTM.Palstra}. 

To sum up, our model provides results is in quantitative agreement with the temperature evolution of entropy, Sommerfeld coefficient and spectral density. Our results are in qualitative agreement with the evolution of Fermi surface, the MSL gap evolution and phase diagram. The extension of our work to the realistic 3D lattice should produce quantitative agreement in (T,P) phase diagram, gap evolution and Fermi surface evolution. Moreoever, it should be possible to calculate realistic electronic band structure for ARPES experiments, Raman scattering and Inelastic Neutron Scattering.
\end{section}

\begin{acknowledgments}
The authors thanks Marie-Aude Measson for helpfull discussion. This research was carried out with the aid of the Computer System of High Performance of the International Institute of Physics – UFRN, Natal, Brazil. This work was also supported, in part, by european IRSES program SIMTECH (contract No. 246937), by CAPES-COFECUB (Brazil-France) and by CNPq (Brazil).
\end{acknowledgments}

\appendix
\begin{section}{Electronic band structure}

\begin{subsection}{General considerations}
The phase diagram presented in section III reveals the existence of several phases characterized by different order parameters. In the present section, we exhibit the specific signature of each of these phases on the electronic band structure and on their associated Fermi surfaces.

We present the electronic band structure in the first Brillouin zone of the square lattice (FBZ) (see figure\ref{fig2}).The electronic band dispersion are plotted along the four directions $M' \Gamma - \Gamma X - XM- M \Gamma$ (see figure\ref{fig2}) where $M', \Gamma, X$ and $M$ are the high symmetry points of the FBZ. Note that the direction $M' \Gamma$ is considered here to emphasize the Z4 symmetry breaking in the MSL phases. 

The zone folding manifest in the AF and MSL phases reduces the First Brillouin zone to the magnetic Brillouin Zone (MBZ) (dotted line on the figure \ref{fig2}). The boundaries of the MBZ are delimited in the four directions $k_{x}=\pm k_{y} \pm \pi$. For simplicity, we signalized the two characteristic points localized on the bound of the MBZ : $X_{1}$ in $(k_{x}=\pi/2,k_{y}=\pi/2)$ and $X_{2}$ in $(k_{x}=-\pi/2,k_{y}=\pi/2)$. 

For each dispersion, we plot the original and the folded electronic band dispersions. The original and folded electronic dispersions are related to the wave vector $Q$. Consequently, an excitation between this two layers indicates the presence of folding and may be related to the excitation wave vector $Q$.

\begin{figure}
  \includegraphics[scale=1]{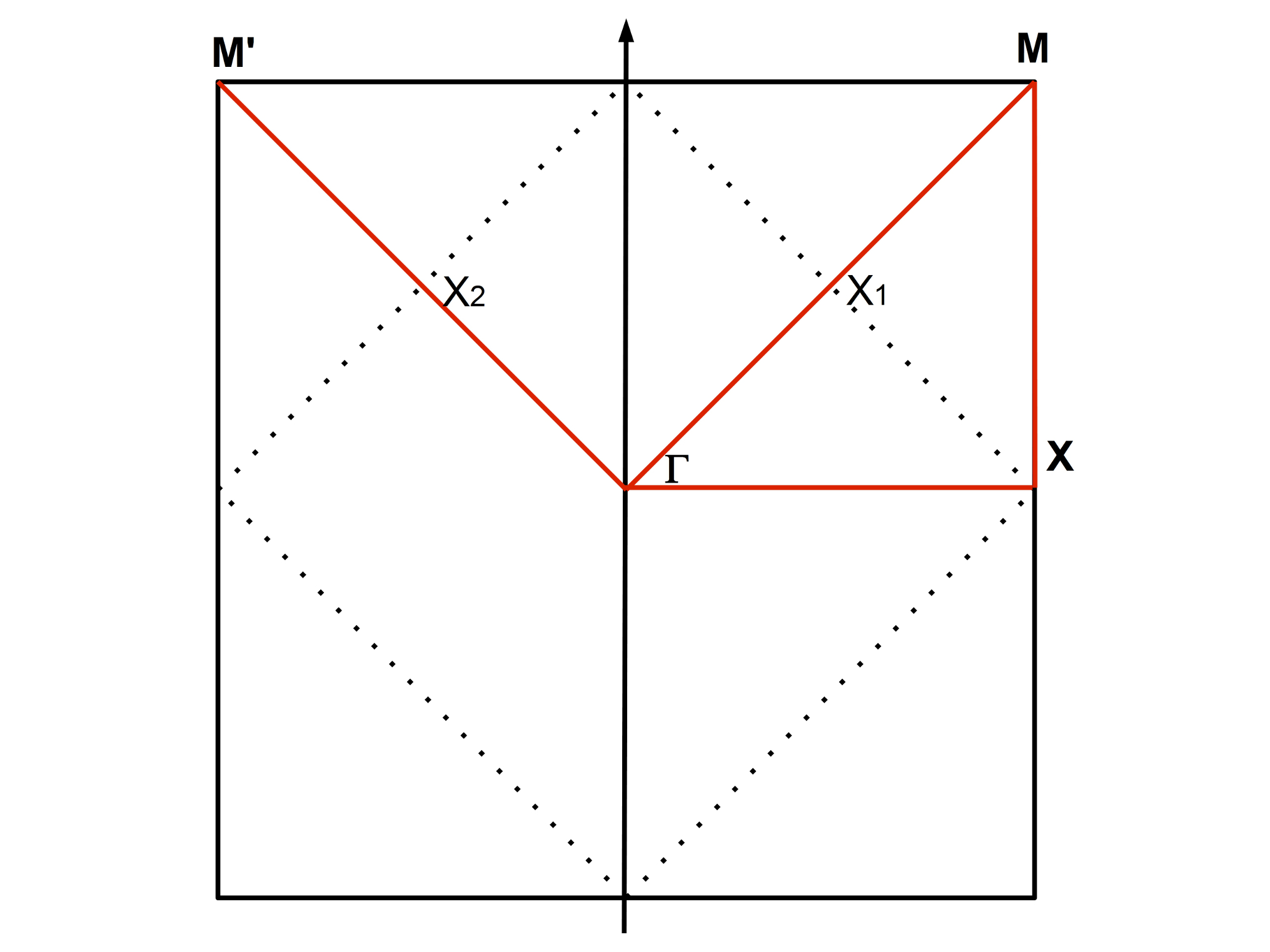}
  \caption{ \label{fig2} The first Brillouin zone of the bidimensionnal square lattice is presented in solid line. For simplicity, the side of the real square lattice is set to $a=1$. The Brillouin zone used in the case of folded Fermi surface is presented in dotted line. The high symmetry point of the first Brillouin zone are the center of the zone $\Gamma$, the center of a side $X$, and a corner $M$ (or $M'$). The characteristical direction of the first Brillouin zone are the directions $\Gamma M$,  $\Gamma X$,  $XM$ and $\Gamma M'$. The $X_{1}$ and $X_{2}$ are two charateristic points of the magnetic Brillouin zone.}
\end{figure}

\end{subsection}

\begin{subsection}{Electronic dispersion in the different phases}

\begin{subsubsection}{The Disordered phases}

\begin{paragraph}{The decoupled phase : Paramagnetic metal and free moments}
~~\\
The electronic band structure of the metal + free moments phase is presented in the figure \ref{fig3}. The spinons are degenerate, without dispersion and centered on zero energy. The spinons are non interacting and non dispersive. 

Note that the energy level of the band crossing (solid and dashed line \ref{fig3} c)) occuring at the $X$, $X_{1}$ and $X_{2}$ points is determined by the chemical potential $\mu_{c}$. These band crossings between the layer and the folded layer occurs in the four directions $k_{y}=\pm \pi \pm k_{x}$. This helps at determine the boundaries of the reduced Brillouin zone (dashed line on figure \ref{fig2}). Nevertheless, the conduction electron dispersion does not present any folding signature, as expected for the metallic phase.

\begin{figure}
  \includegraphics[scale=1]{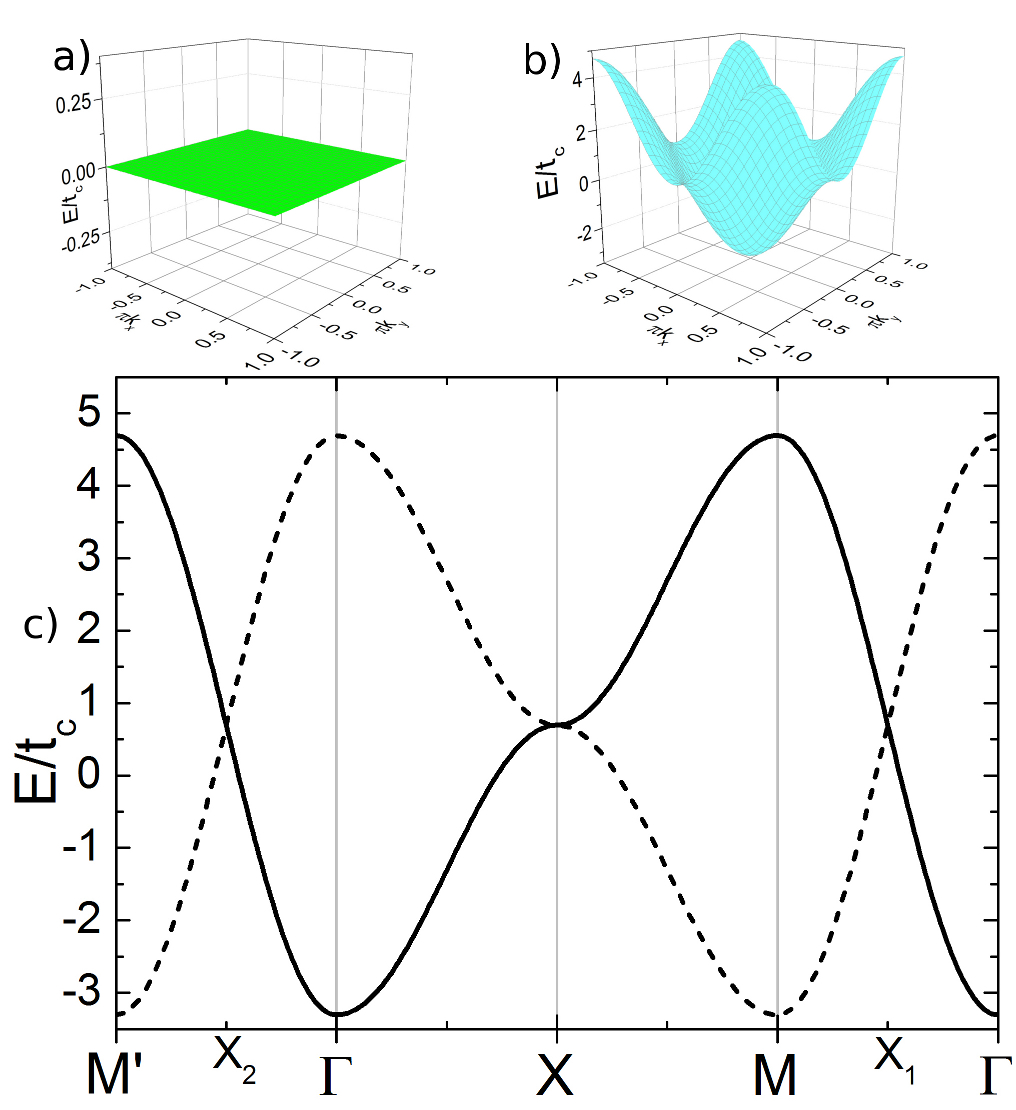}
  \caption{ \label{fig3} The electronic band structure of the local moments a) and the conduction electons b) in the paramagnetic metal + free moments phase in the first Brillouin zone. In this case $\phi_{0}=0$, $\phi_{Q}=0$, $\rho=0$, $\theta=0$, $S_{Q}=0$, $\mu=0$ and $\mu_{c}/T_{K0}=-7.45$ at $T/T_{K0}=2.$, $J_{SL}=0$ and $J_{AF}=0$. In the figure c) are plotted the original (solid) and folded (dashed) dispersions of the c electron in the characteristic directions of the first Brilouin zone. The Fermi level is set at E=0.}
\end{figure}

\end{paragraph}
\begin{paragraph}{The Kondo phase with $\phi_{0}=0$}
~~\\
The electronic band structure and the electronic dispersion of the Kondo phase with $\phi_{0}=0$  are plotted in the figure \ref{fig6}. There is now the appearence of a gap in the electronic band structure around the Fermi level. The hybridization between the conduction electrons and the local moments leads to an extended Fermi surface (FS). This FS enlargement is associated with the presence of the Kondo quasiparticles in the system and is related to the decreasing of the resistivity occuring below $T_{K}$ in Kondo lattice systems \cite{ACHewson}.
This indirect gap induced by Kondo interaction (see figure \ref{fig6} b)) approximatively equal to $ \frac{1}{2}(t_{c}-\sqrt(t_{c}^{2}+4\rho^{2}))$ depending on the electronic bandwidth and on the amplitude of the Kondo coupling $\rho$ and vanishes with the Kondo phases.

\begin{figure}
  \includegraphics[scale=1]{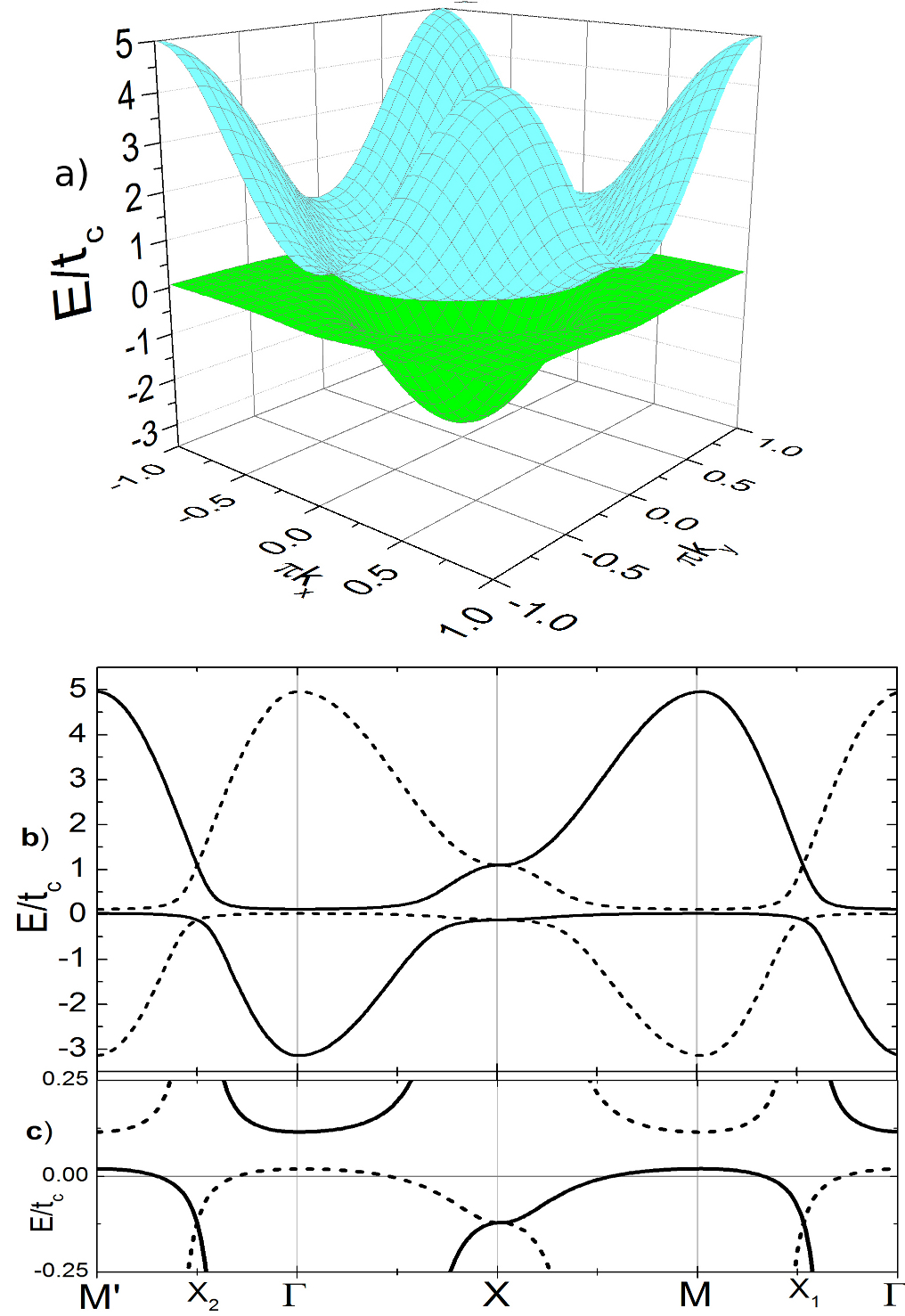}
  \caption{ \label{fig6} In a), the electronic band structure of the Kondo phase  with $\phi_{0}=0$ in the first Brillouin zone. In this case $\phi_{0}=0$, $\phi_{Q}=0$, $\rho/T_{K0}=4.40$, $\theta=0$, $S_{Q}=0$, $\mu/T_{K0}=-0.58$ and $\mu_{c}/T_{K0}=-9.37$ with $T/T_{K0}=0.5$, $J_{SL}=0$ and $J_{AF}=0$. In the figure b) is plotted the original (solid) and folded (dashed) dispersions of the electronic band in the four characteristic directions of the first Brilouin zone. The zoom near the Fermi level (E=0) is shown in the figure c). Note the direct gap in the $\Gamma$ and $M$ point.}
\end{figure}
\end{paragraph}

\begin{paragraph}{The Kondo phase with $\phi_{0} \neq 0$}
~~\\
The electronic band structure and the single particle dispersion of the Kondo phase with $\phi_{0} \neq 0$ are plotted on the figure \ref{fig7}. In this phase, the Kondo screening hybridizes two layers with opposite dispersions. This implies the opening of  a direct gap between the two layers. The direct gap at the $\Gamma$ point is proportionnal  to $\rho^{2}/4t_{c}$. This gap vanishes with the Kondo phase. Note that no symmetry is broken between the $\phi=0$ and $\phi \neq 0$ Kondo phases. This is typical of a crossover separating these two phases. Also, this crossover is accompagnied by a change of the Fermi surface (see fig \ref{fig122} d) and e)) .

\begin{figure}
  \includegraphics[scale=1]{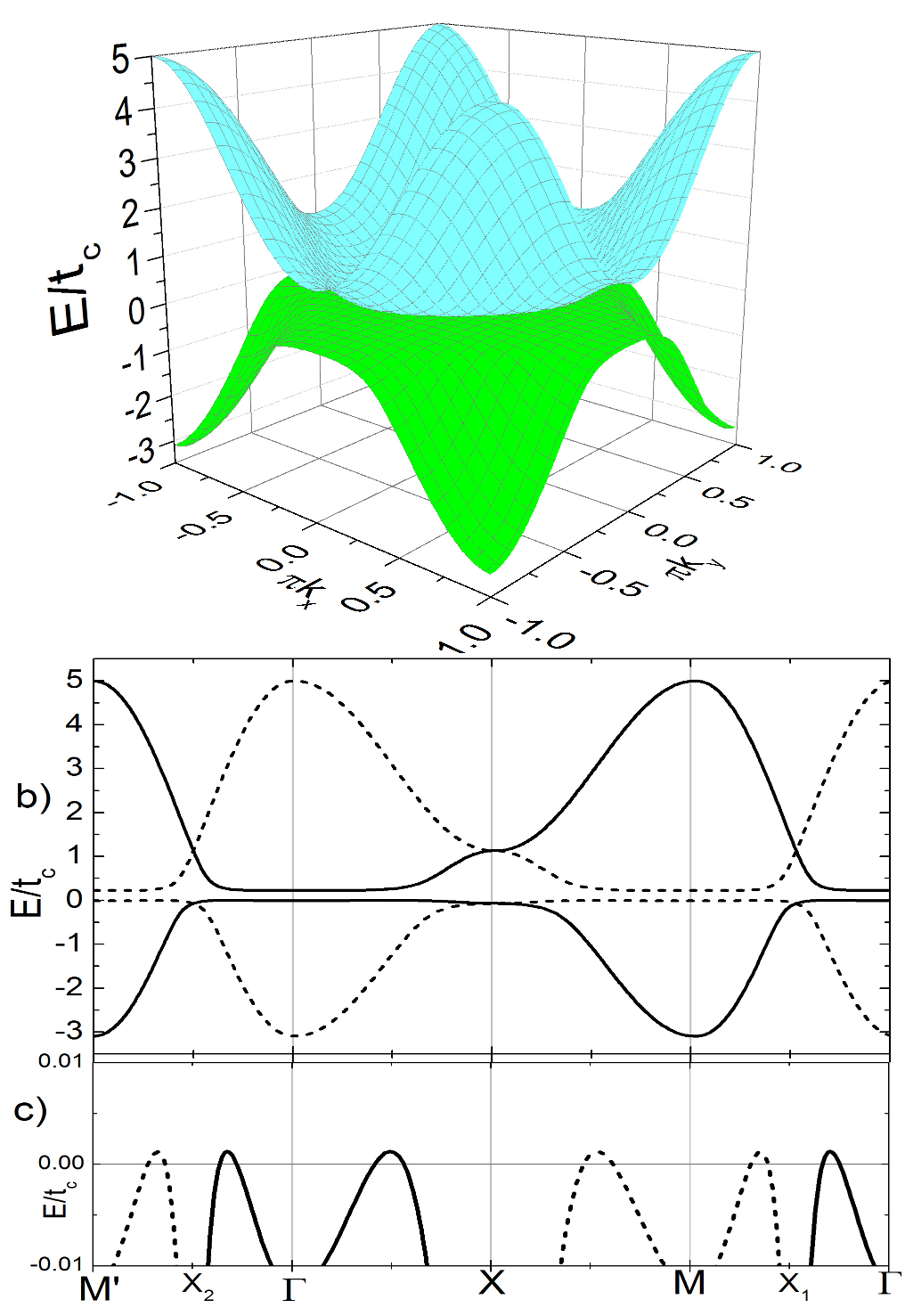}
  \caption{ \label{fig7}  In a), the electronic band structure of the Kondo phase $\phi_{0} \neq 0$ in the first Brillouin zone.  In this case $\phi_{0}/T_{K0}=0.19$, $\phi_{Q}=0$, $\rho/T_{K0}=4.25$, $\theta=-\pi/2$, $S_{Q}=0.$, $\mu/T_{K0}=-0.99$ and $\mu_{c}/T_{K0}=-9.81$ with $T/T_{K0}=0.5$, $J_{SL}/T_{K0}=0.5$ and $J_{AF}=0$. In the figure b) is plotted the original (solid) and folded (dashed) dispersions of the electronic band in the four characteristic directions of the first Brilouin zone. The zoom near the Fermi level (E=0) is shown in the figure c). Note the direct gap in the $\Gamma$ and $M$ point.}
\end{figure}
\end{paragraph}

\end{subsubsection}

\begin{subsubsection}{The decoulped phase : AF+metal}
In the AF phase, the electronic band structure is composed by the local moments (see figure \ref{fig4} a)) and the conduction electrons (see figure \ref{fig4} b)).  In our model, the spinons in the AF phase are totally flat and non degenerate (see figure \ref{fig4} a)). The gap between the two spinons is equals to $8.S_{Q}$. 

From the magnetic type Kondo mean field decoupling \cite{C.Lacroix}, we deduce the magnetic interaction of the local moments with the conduction electrons. Considering the first order perturbation of the Kondo interaction in the AF phase, the mean field Kondo lattice Hamiltonian writes $H^{MF}_{KL} \approx J_{K}\sum_{q,\sigma}\left[c_{q\sigma}^{+}(\sigma S_{Q})c_{q-Q\sigma}\right] $. The conducting electron spectrum writes $E^{c} \approx \frac{1}{2}t_{c}(\epsilon_{q}+\epsilon_{q-Q})-\lambda \pm  \frac{1}{2}\sqrt{t_{c}^{2}(\epsilon_{q-Q}-\epsilon_{q})^{2}+4S_{Q}} $. The magnetization of the local moments induce a gap in the conduction electrons band structure (see figure \ref{fig4} c) and d)). The direct gap  at the $X$, $X_{1}$ and $X_{2}$ points between the original (solid) and the folded (dashed) layer equals $2S_{Q}$ (see figure \ref{fig4}). The magnetization seen by the conduction electrons is four times smaller than for the corresponding local moments because of the local aspect of the Kondo coupling. This gap opening is accompanied by a folding of the conduction electron band structure in relation with the lattice symmetry breaking. Moreover, this gap is related to the commensurable vector $Q$.

\begin{figure}
  \includegraphics[scale=1]{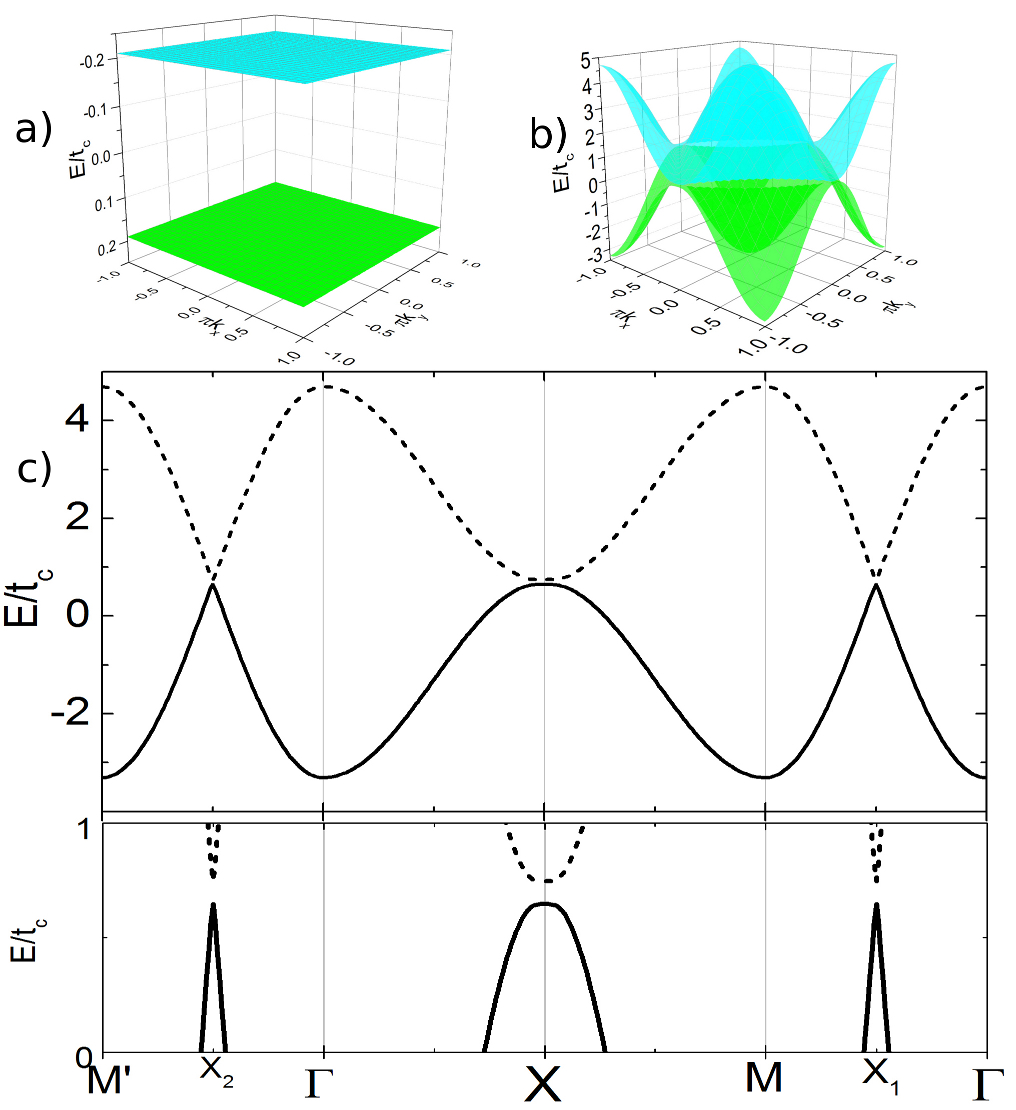}
  \caption{ \label{fig4} In a), the electronic band structure of the the f electrons and the conducting electrons b) in the AF phase in the first Brillouin zone. In the AF state $\phi_{0}=0$, $\phi_{Q}=0$, $\rho=0$, $\theta=0$, $S_{Q}/T_{K0}=0.51$, $\mu/T_{K0}=0.10$ and $\mu_{c}=-7.10$ at $T/T_{K0}=0.01$, $J_{SL}=0.$ and $J_{AF}/T_{K0}=0.5$. In the figure c) are plotted the original (solid) and folded (dashed) dispersions of the c electron band in the characteristic directions of the first Brilouin zone. The zoom near the Fermi level (E=0) is shown in the figure d).}
\end{figure}
\end{subsubsection}

\begin{subsubsection}{The MSL phases}
\begin{paragraph}{The Decoupled phase : MSL + metal}
~~\\
The MSL phase is an itinerant phase associated with the dispersion relation $\gamma_{q-\frac{Q}{2}}$ whose dispersion is presented in Figure \ref{fig5}. We clearly see a Z4 symmetry breaking with the opening of the direct gap at the point $X_{1}$, called $\Delta_{X_{1}}$, absent at the $X_{2}$ point (see figure \ref{fig5} c)). This Z4 symmetry breaking has been introduced in the initial MSL model \cite{C.Pepin} following the observation on susceptibility measurements \cite{ROkazaki} in the HO phase of $URu_{2}Si_{2}$ compound. At the $X_{2}$ point, the two bands are in contact with the Fermi level. In fact, this contact exists along the flat bands directions $k_{y}=\pm \pi +k_{x}$. Note that hole or electron pockets emerge along these two directions in the spinon dispersion at the conduction electron Fermi surface in the presence of second nearest neighbour term in the dispersion relation \cite{C.Pepin}. The gap at the $X_{1}$ point, $\Delta_{X_{1}}=8\phi_{Q}$, can be associated with the vector $Q$ and only depend on the modulated spin liquid mean field parameters The spinon dispersion reveals a non-symetrical folding, originating in the association of N\'{e}el order lattice symmetry breaking with Z4 symmetry breaking.

\begin{figure}
  \includegraphics[scale=1]{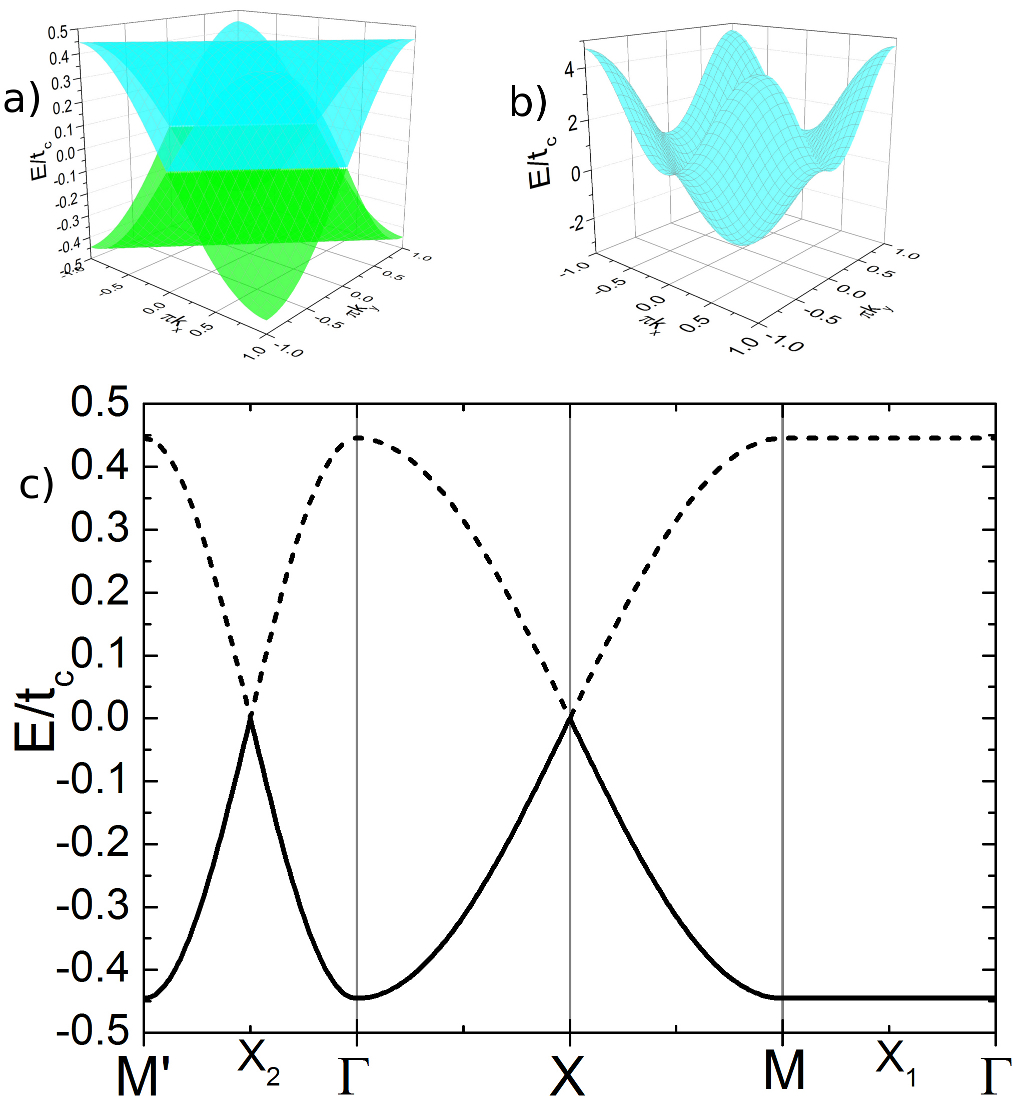}
  \caption{ \label{fig5} In a), the electronic band structure of the the f electrons and the conducting electrons b) in the MSL phase in the first Brillouin zone. In the MSL state $\phi_{0}/T_{K0}=1.14$, $\phi_{Q}/T_{K0}=1.14$, $\rho=0$, $\theta=0$, $S_{Q}=0$, $\mu=0$ and $\mu_{c}/T_{K0}=-7.10$ with $T/T_{K0}=0.01$, $J_{SL}/T_{K0}=5$ and $J_{AF}=0$. In the figure c) are plottedthe original (solid) and folded (dashed) dispersions of the c electron band in the characteristic directions of the first Brilouin zone. The Fermi level is set at $E=0$.}
\end{figure}

\end{paragraph}
\begin{paragraph}{The Correlated phase :KMSL}
~~\\
The transformation of the electronic band structure at the Kondo-KMSL phase transition can be observed throughout the figure \ref{fig7} and \ref{fig8}. We observe a strong electronic band reconstruction with the emergence of folding due to the lattice symmetry breaking. Moreover, the opening of a gap only at the $X_{1}$ point emphasizes the Z4 symmetry breaking in the KMSL phase. This gap is also associated with the commensurate vector $Q$. The Z4 symmetry breaking is related to the modulated spin liquid mean field $\phi_{Q}$. The direct gap occuring at the $X_{1}$ point,$\Delta_{X_{1}}$, is approximatively equal to $4 \phi_{Q}$. This gap is a characteristic signature of the MSL phase.

\begin{figure}
  \includegraphics[scale=1]{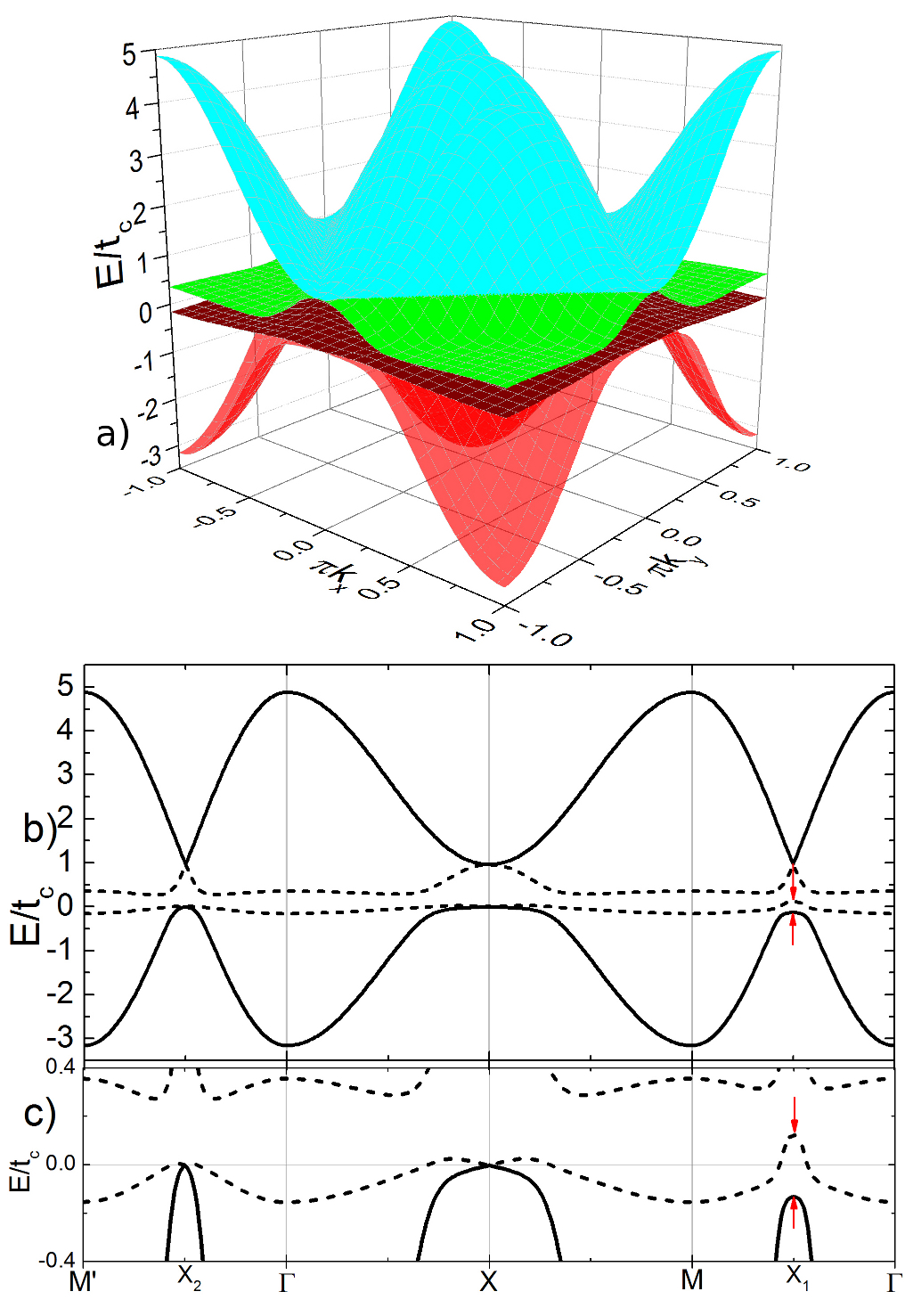}
  \caption{ \label{fig8} In a), the electronic band structure of the KMSL phase in the first Brillouin zone. In this case $\phi_{0}/T_{K0}=0.6$, $\phi_{Q}/T_{K0}=0.36$, $\rho/T_{K0}=2.99$, $\theta=-\pi/2$, $S_{Q}=0$, $\mu/T_{K0}=-0.98$ and $\mu_{c}/T_{K0}=-8.8$ with $T/T_{K0}=0.01$, $J_{SL}/T_{K0}=1.84$ and $J_{AF}/T_{K0}=0.05$ . In the figure b) is plotted the original (solid) and folded (dashed) dispersions of the electronic band in the four characteristic directions of the first Brilouin zone. The zoom near the Fermi level (E=0) is shown in the figure c). Note the direct gap in the $\Gamma$ and $M$ point and the gap at the $X_{1}$ point.}
\end{figure}
\end{paragraph}
\end{subsubsection}

\begin{subsubsection}{The coexisting phase}
 The dispersion of the electronic band structure of the coexisting KMSL-AF state is presented in figure \ref{fig9}. The gap induced by the modulated spin liquid is compensated by the AF gap. Consequently, the Z4 symmetry breaking is progressively destroyed. On the electronic band dispersion shown in Figure \ref{fig9} c), we see the onset of a gap between the original and folded electronic band that does not exist in the pure KMSL phase electronic dispersion (see figure \ref{fig8} c)). This direct gap at the $X_{2}$ and the X points ($k_{x}=\pi$,$k_{y}=0$) approximatively equals to $4 S_{Q}$. It is directely related to the magnetization of the system. This gap exists only in the antiferromagnetic phase and implies a less pronounced aspect of the Z4 symmetry of the Fermi surface. The direct gap $\Delta_{X_{1}}$ is approximatively equal to $4\sqrt{\phi_{Q}^{2}+S_{Q}^{2}}$ and it also depends on the magnetization of the systems. Note that both direct gaps $\Delta_{X_{1}}$ and $\Delta_{X_{2}}$ are related to the wave vector $Q$ and this emphasizes the similar symmetry breaking between the AF and MSL phases.

\begin{figure}
  \includegraphics[scale=1]{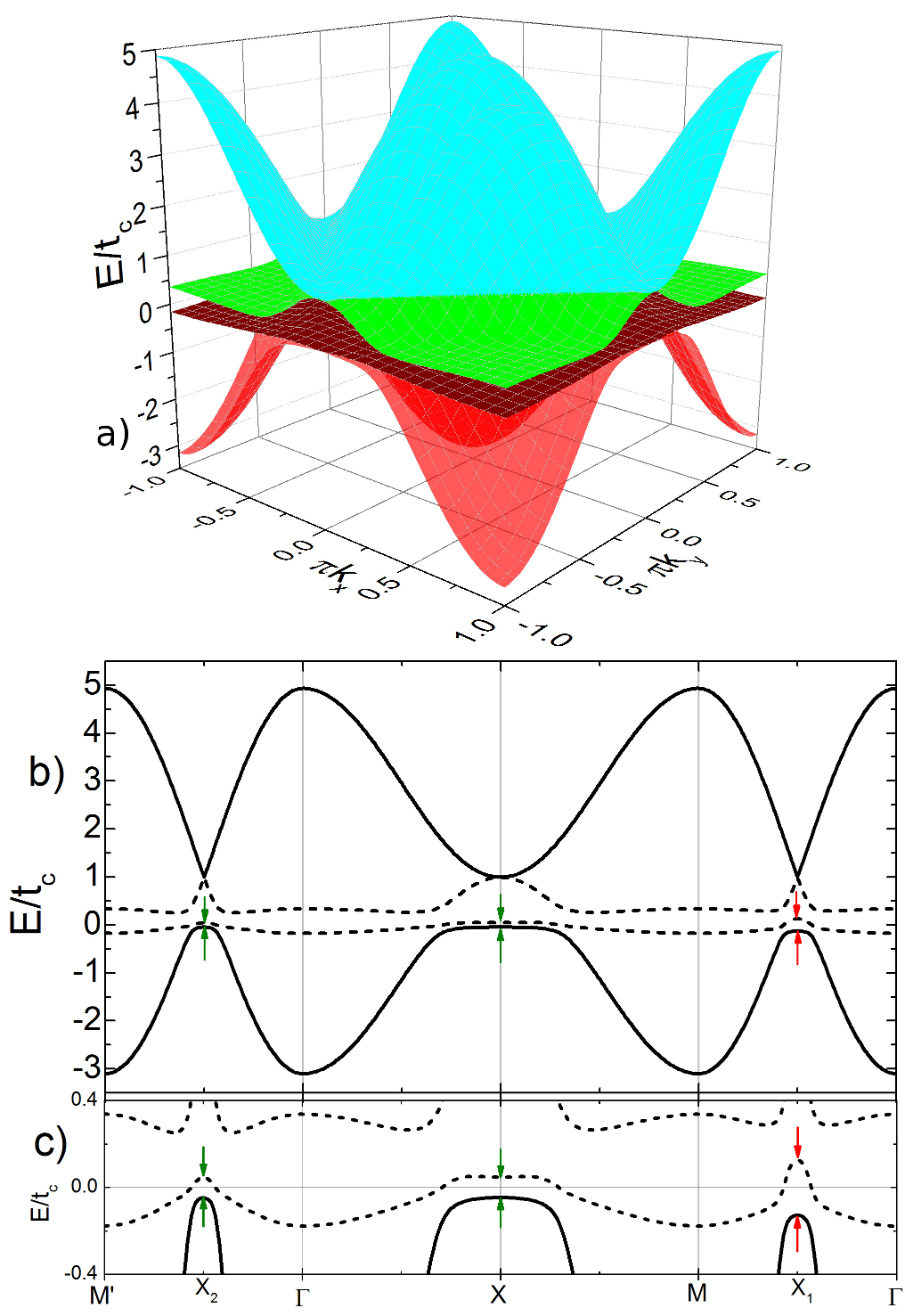}
  \caption{ \label{fig9} In a), the electronic band structure of the KMSL-AF phase in the first Brillouin zone. In this case $\phi_{0}/T_{K0}=0.6$, $\phi_{Q}/T_{K0}=0.33$, $\rho/T_{K0}=2.7$, $\theta=-\pi/2$, $S_{Q}/T_{K0}=0.13$, $\mu/T_{K0}=-0.78$ and $\mu_{c}/T_{K0}=-8,35$  with $T/T_{K0}=0.01$, $J_{SL}/T_{K0}=1.84$ and $J_{AF}/T_{K0}=0.15$ . In the figure b) is plotted the original (solid) and folded (dashed) dispersions of the electronic band in the four characteristic directions of the first Brilouin zone. The zoom near the Fermi level (E=0) is shown in the figure c).}
\end{figure}
\end{subsubsection}
\end{subsection}
\end{section}

\begin{section}{Fermi Surfaces: Characteristical signatures of the different phases}

We display the evolution of the electronic band structure between the different phases. We observe the onset of gaps the systems characterizing the existing different phases. Moreover, we see that the electronic band structure evolves in the different phases, as shown in the figure \ref{12}.The system can exhibit four types of order parameters that can coexist in the system.

\begin{subsection}{The Kondo order parameter}
 The Kondo effect is characterized by a volume change of the Fermi surface. This change of volume is a consequence of the hybridization of the f and c electrons.  It occurs between the decoupled phases (see \ref{12} a), b) and f)) and the Kondo phase (figure \ref{fig122} d) and e)).  This explains the Kondo breakdown and the accompanied existence of the QCP in the phase diagram at $J^{\star}_{SL}$ and $J^{\star c}_{AF}$ (see fig \ref{fig99}). The Kondo breakdown is difficult to observe at the KMSL to AF phase transition because of the lattice symmetry breaking occuring at these phase transition. Nevertheless, change of Fermi surface volume, without folding, is a clear signature of the Kondo effect.
\end{subsection}

\begin{subsection}{The lattice symmetry breaking}
 The lattice symmetry breaking occuring in the magnetic ordered phases manifests itself in the folding (or non-symmetrical folding) of the Fermi surface (see  (see \ref{fig122} b) c) f) and g). These lattice symmetry breaking is a clear signature of the magnetically ordered nature of the phases. The folding is observable by the periodicity which shows up in the Fermi surface. We see that this periodicity is related to the limit of the MZB. This folding can also be observed by neutron scattering. The folding of the Fermi surface is a typical feature of magnetically ordered phases (AF and MSL). The consequence of that is the enlargement of the Fermi surface which is not related to Kondo breakdown physics in this case.
\end{subsection}

\begin{subsection}{The Z4 symmetry breaking}
 These symmetry is broken in the Fermi surfaces with MSL charateristics  (see \ref{fig122} d),c) and g)). The Z4 symmetry breaking is emphasized by the direct gap opening in the electronic dispersion  along the flat band direction $k_{y}=-k_{x} \pm \pi$  which is observable at $X_{1}$  on the figures \ref{fig6}, \ref{fig8} and \ref{fig9} . The Z4 symmetry breaking of the Fermi surface is a characteristic signature of the MSL states.
\end{subsection}

\begin{subsection}{The time reversal symmetry breaking}
 The time reversal symmetry occurs only in the AF phase and originates the AF order parameter. These symmetry cannot be directly seen on the Fermi surface. However, the electronic dispersion associated with the AF phase (figure \ref{fig4} and \ref{fig9}) exhibits a gap opening between the two conduction electron bands. These specific gap, proportionnal to $S_{Q}$ and related to the commensurate vector $Q$, could be a specific signature of the time reversal symmetry.
\end{subsection}

The difference between the Kondo phases with $\phi_{0} =0 $ and $\phi_{0} \neq 0$ is observable in the electronic band structure and in the corresponding Fermi surfaces. If the Fermi surfaces of these two phase are different, they preserve the same symmetry and conserve the same volume. A crossover should describes the transition between these two phases.
 A gauge invariance manifest itself in the Hamiltonian. We see that the Hamiltonian is invariant under the transformation $(\phi_{0},\phi_{Q},b_{0},ib_{Q}) \longrightarrow (-\phi_{0},-\phi_{Q},ib_{Q},b_{0})$. These invariance corresponds to the invariance by changing the phase of the fermions $f$ like $f_{R,\sigma}\longrightarrow e^{i(\frac{\pi}{2}-Q.R)}f_{R,\sigma}$. These invariance gives another explanation of the Kondo phase with $\phi_{0} \neq 0$. The inhomogeneous coupling between the two phases can be understood as a homogeneous Kondo coupling between two layers with opposite dispersions. These homogeneous coupling does not change the symmetry of the phases.

\end{section}

\end{document}